\documentclass[12pt,twoside,onecolumn,draft]{IEEEtran}

\usepackage{epsfig}
\usepackage{graphicx,cite,amssymb,amsmath,pstricks}

\usepackage[printonlyused]{acronym}

\newcommand{\PX}[1] {{\mathbb{P}}\left\{{#1}\right\}}
\newcommand{\EX}[1] {{\mathbb{E}}\left\{{#1}\right\}}

\def\erfc{\text{erfc}}

\def\rank{\text{rank}}

\begin{document}             % End of preamble and beginning of text.

\begin{acronym} % usage: \ac{SW}, \acp{SW} for plurals \acf{SW} Use the full name of the acronym. \acs{SW}Use the acronym, even before the first corresponding \ac command \acl{acronym}Expand the acronym without using the acronym itself.
\acro{SW}{sync word} \acro{FS}{frame synchronization}
\acro{r.v.}{random variable} \acro{i.i.d.}{independent,
identically distributed} \acro{p.d.f.}{probability distribution
function} \acro{c.d.f.}{cumulative distribution function}
\acro{ch.f.}{characteristic function} \acro{AWGN}{additive white
gaussian noise} \acro{BSC}{binary symmetric channels}
\acro{BPSK}{binary phase shift keying} \acro{SNR}{signal-to-noise
ratio} \acro{STC}{space-time codes} \acro{P-STC}{pragmatic
space-time codes}\acro{BFC}{block fading channels}
\acro{CC}{convolutional codes} \acro{PEP}{pairwise error
probability} \acro{QPSK}{quaternary phase shift keying}
\acro{M-QAM}{M-ary quadrature amplitude modulation}
\acro{GTF}{generalized transfer function} \acro{ST-GTF}{space-time
generalized transfer function} \acro{FER}{frame error rate}
\acro{BER}{bit error rate} \acro{ML}{maximum likelihood}
\acro{MIMO}{multiple input multiple output} \acro{RBFC}{reference block fading channel}
\end{acronym}

\markboth{Submitted to IEEE Transactions on Information Theory%,
%Vol. XX, No. Y, Month 2002
}{M.Chiani, A.Conti, V.Tralli:
 Pragmatic  Space-Time Trellis Codes ...}

\title{\mbox{}\vspace{0.5cm}\\
    Pragmatic Space-Time Trellis Codes for Block Fading Channels
    \vspace{1.0cm}}

\author{$\mbox{Marco~Chiani}^{}$,~\IEEEmembership{Senior
Member,~IEEE,}
$\mbox{Andrea~Conti}^{}$,~\IEEEmembership{Member,~IEEE} and
$\mbox{Velio~Tralli}^{}$,~\IEEEmembership{Senior Member,~IEEE,}
%\thanks{${}^{\dag}$ Corresponding author.}
%\thanks{This paper was presented in part at ICC 2001, Helsinki, June 2001.}
%\thanks{This research was supported by Ministero dell'Istruzione, Universit$\grave{\text{a}}$ e della Ricerca Scientifica (MIUR), ...}
\thanks{Marco Chiani
    is with DEIS, University of Bologna,
    40136 Bologna, ITALY
    (e-mail: {\tt mchiani@deis.unibo.it}). Andrea Conti and Velio Tralli are with ENDIF,
University of Ferrara, and WiLab, Universiy of Bologna, Italy, (e-mail:  {\tt a.conti@ieee.org}, {\tt
vtralli@ing.unife.it}).}
\\
\vspace{0.5cm}
\underline{Corresponding Address:}\\
    Marco Chiani\\
    DEIS, University of Bologna\\
    V.le Risorgimento 2, 40136 Bologna, ITALY\\
\bigskip
    Tel: +39-0512093084 \qquad Fax: +39-0512093540\\
    e-mail: {\tt mchiani@deis.unibo.it}
}

%\date{To be submitted to IEEE Transactions on Communications, \today.}
\date{%Revised
\today.}
%\date{\empty}

%\thispagestyle{empty}
%\pagestyle{empty}
\setcounter{page}{100}
\maketitle

%\renewcommand{\baselinestretch}{1.55} \normalsize
%\renewcommand{\baselinestretch}{1.65} \normalsize
% The above \baselinestretch command does not change automatically.
% A font size changing command should be executed to make the new value
% in effect.  I trick this by using \normalsize, which does not change
% the font size!
%%%%%%%%%%%%%%%%%%%%%%%%%%%%%%%%%%%%%%%%%%%%%%%%%%%%%%%%%%%%%%%%%%%%%%
\newpage

\begin{abstract}
A pragmatic approach for the construction of \acl{STC} over
\acl{BFC} is investigated. The approach consists in using common
convolutional encoders and Viterbi decoders with suitable generators
and rates, thus greatly simplifying the implementation of \acl{STC}.

For the design of \acl{P-STC} a methodology is proposed and applied, based on the extension of the concept of \acl{GTF} for \acl{CC} over
\acl{BFC}. Our search algorithm produces the convolutional encoder generators
of  %optimum, with respect to our criterion,
\acl{P-STC} for various number of states, number of antennas and
fading rate.

Finally it is shown that, for the investigated cases, the performance
of \acl{P-STC} is better than that of previously known \acl{STC},
confirming that they are a valuable choice in terms of both
implementation complexity and performance.

%
%It is shown that these \acl{P-STC} are flexible and achieve
%excellent performance both on the quasi-static fading channel and in
%general for different fading rates, taken into account by the
%\acl{BFC} model.

%A pragmatic approach to space-time codes (STC) over a block fading
%channel (\ac{BFC}) is addressed. It consists in using common
%convolutional codes to obtain STC, greatly simplifying the encoder
%and the decoder. It is shown that pragmatic STC (\ac{P-STC}) achieve a
%good performance, similar to that of the best known STC, and that
%they are suitable for systems with different spectral efficiencies
%and fading velocity (taken into account by the \ac{BFC} model). To
%design \ac{P-STC} for \ac{BPSK} and QPSK modulation formats we propose a
%search algorithm based on a proper formulation of the pairwise
%error probability and the error enumerating function for
%wide-sense geometrically uniform STC over \ac{BFC}. Comparisons with
%other STC proposals are also provided, showing that this \ac{P-STC}
%scheme is a good compromise between complexity and performance.
\end{abstract}


%%%%%%%%%%%%%%%%%%%%%%%%%%%%%%%%%%%%%%%%%%%%%%%%%%%%%%%%%%%%%%%%%%%%%%%%%%%%%%%

\section{Introduction}
%\note{MARCO}
%In radio communication the received power level fluctuates due to multipath propagation.
It is known since many years that the use of multiple receiving
antennas, sufficiently spaced apart each other to obtain
independent copies of the transmitted signal, is an efficient way
to mitigate the effects of multipath propagation
(see, e.g., \cite{SchBenSte:B96,Pro:B01,Wint:98a}).
%That based on this approach are called (receiving) antenna diversity techniques.
However, only recently it has been realized that even the use of
multiple transmitting antennas can give similar improvements
\cite{Ala:98,NarTroWor:99,Wit:93}.
%In this regard, an
%information theoretic study of transmit diversity techniques is
%given in \cite{NarTroWor:99}.
With the introduction of \ac{STC} it has been shown how, with the
use of proper trellis codes, multiple transmitting antennas can be
exploited to improve system performance obtaining both diversity
and coding gain, without sacrificing spectral efficiency
\cite{Wit:93, GueFitBelKuo:96,
TarSesCal:98,TarNagSesCal:99,VucYua:B03,Jaf:B05}.

In particular, the design of \ac{STC} over quasi-static flat
fading (i.e., fading level constant over a frame and independent
frame by frame) has been addressed in \cite{TarSesCal:98}, where
some handcrafted trellis codes for two transmitting antennas have
been proposed. A number of extensions of this work have eventually
appeared in the literature to design good codes for different
scenarios, and \ac{STC} with improved coding gain have been
presented in \cite{BarBauHan:00, CheYuaVuc:01,YanBlu:02}.
In \cite{HamElg:00} it is pointed out that the diversity achievable
by \ac{STC} for \ac{BPSK} and \ac{QPSK} modulations can also be
investigated by a binary design criteria, instead of looking for
the %Euclidean
distances among complex transmitted sequences. This approach has been extended to \ac{MIMO} \ac{BFC} in \cite{ElGHam:03}.

The determination of the \ac{STC} with maximum diversity
gain and largest coding gain remains a difficult task, especially
for a large number of transmitting antennas and %large number of
trellis code states. Moreover, the design of \ac{STC} for fast
fading channels is still an open problem.

In this paper we present an approach to \ac{STC} to
simplify the encoder and decoder structures, that also allows a
feasible method to search for good codes in \ac{BFC} \cite{MceSta:84, Chi:J98, MalLei:99}. In fact, a
criterion to achieve maximum diversity is given in
\cite{TarSesCal:98}, where, however, coding gain optimization is not
addressed. Moreover, the \ac{STC} in \cite{TarSesCal:98} require
ad-hoc encoders and decoders.
For these reasons, %and as one of the principal contribution of this paper,
we present another possible approach to space-time coding,
denominated \ac{P-STC} \cite{ChiConTra:C01a}. Here, the
``pragmatic'' approach (the name following \cite{VitWolZehPad:89})
consists in the use of common convolutional encoders and Viterbi
decoders over multiple transmitting and receiving antennas. We
show that \ac{P-STC} achieve maximum diversity and excellent
performance, with no need of specific encoder or decoder different
from those used for \ac{CC}; the Viterbi decoder requires only a
simple modification in the metrics computation.

We use the \ac{BFC} model to investigate the design and the
performance of \ac{STC}. The
\ac{BFC} represents a simple and powerful model to include a variety
of fading rates, from "fast" fading (i.e., ideal symbol
interleaving) to quasi-static.

Here, after the proposal of the \ac{P-STC} structure, we first
derive the \ac{PEP} of \ac{STC} over block fading channels. Then,
we propose a method based on suitable error trellis diagrams and
generalized transfer function to evaluate a bound on the
performance of
%geometrically uniform (GU)
\ac{STC} over \ac{BFC}, with a discussion on geometrical
uniformity over the \ac{BFC}.
%We apply the same
%methodology to a wider class of STC that we denote as wide-sense
%GU.

 A new algorithm for searching good \ac{P-STC} over \ac{BFC} is then
presented and applied to obtain the optimum (with respect to our
performance bound) convolutional generators for various constraint
lengths and fading rates. The numerical results, which compares
our \ac{P-STC} with the best known \ac{STC}, confirm the validity
of the approach.

For simplicity we will focus on the \ac{BPSK} and \ac{QPSK}
modulation formats, but the extension to other formats such as
\ac{M-QAM} is straightforward.

The paper is organized as follows: in section II the channel model
and the general architecture of a system with \ac{STC} are
described; in section III the \ac{P-STC} are presented; in section
IV the \ac{PEP} for \ac{STC} over \ac{BFC} is derived; in section
V the frame error probability for \ac{STC} over \ac{BFC} is
analyzed; in section VI the search methodology for \ac{P-STC} in
\ac{BFC} is illustrated; in section VII numerical results are
provided, followed by the conclusions in section VIII.
% geometrically
%uniform STC are investigated, with an algorithm which evaluates
%the modified weight enumerating function. The class of Pragmatic
%STC is proposed in section V and, finally, some numerical results
%are given in section VI.

%\note{VERIFICARE PAPERI PIU' RECENTI}

%%%%%%%%%%%%%%%%%%%%%%%%%%%%%%%%%%%%%%%%%%%%%%%%%%%%%%%%
\section{System architecture and channel model}
%%%%%%%%%%%%%%%%%%%%%%%%%%%%%%%%%%%%%%%%%%%%%%%%%%%%%%%%%%%%%%%%%%%%%%%%%%%%%%%%
%\note{MARCO}

\begin{figure}
%\centerline{\includegraphics[width=0.95\linewidth,draft=false]
%    {schemePSTC.eps}}
{\tiny \centerline{\input{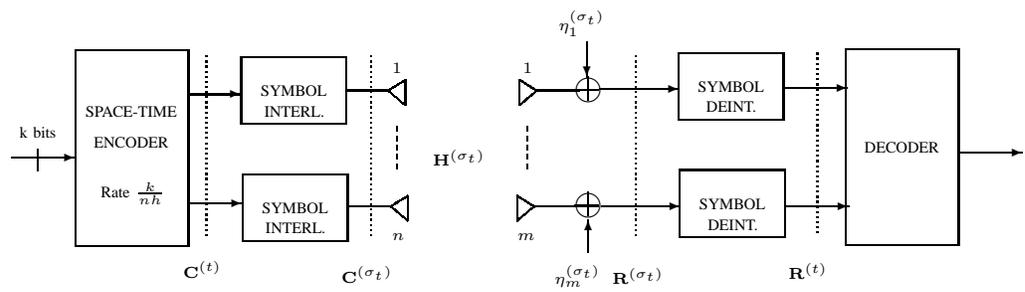}}} \vskip -8cm
\caption{Equivalent low-pass scheme for space-time codes.}
\label{fig:schemeSTC}
\end{figure}

The general low-pass equivalent scheme for space time codes is
depicted in Fig.~\ref{fig:schemeSTC}, where $n$ and $m$ denote the
number of transmitting and receiving antennas, respectively. %It includes a space-time
%encoder, an Interleaver (I), a space time modulator, a fading channel (C), a space-time
%demodulator, a deinterleaver (DI) and a soft-decision space-time Viterbi decoder.

We indicate\footnote{The superscripts $^{H}$, $^{T}$ and $^{*}$
denote conjugation and transposition, transposition only, and
conjugation only, respectively.}
 with ${\bf C}^{(t)}=\left[ c_{1}^{(t)} , \ldots, c_{n}^{(t)}\right] ^{T}$
 a super-symbol, that is a vector of symbols simultaneously transmitted at
 discrete time $t$ on the $n$ antennas,
each having unitary norm and generated according to the modulation
format by proper mapping. Thus, $n$ symbols are sent in parallel
on the $n$ transmitting antennas. A codeword is a sequence
$\underline{c}=\left({\bf C}^{(1)}, \ldots,{\bf
C}^{(N)}\right)$ of $N$ super-symbols generated by the encoder. % for each symbol interval.
%$$\underline{c}=c_{1}^{(1)},c_{2}^{(1)},\cdot \cdot \cdot
%,c_{n}^{(1)},c_{1}^{(2)},c_{2}^{(2)},\cdot \cdot \cdot ,c_{n}^{(2)},\cdot
%\cdot \cdot c_{1}^{(N)},c_{2}^{(N)},\cdot \cdot \cdot c_{n}^{(N)}.$$

This codeword $\underline{c}$ is first interleaved (we refer to
intra-codeword interleaving) to obtain the sequence
$\underline{c}_I={\mathcal{I}}(\underline{c})=\left({\bf
C}^{(\sigma_1)}, \ldots,{\bf C}^{(\sigma_N)}\right)$, where
$\sigma_1,\ldots,\sigma_N$ is a permutation of the integers $1,
\ldots, N$ and ${\mathcal{I}}(\cdot)$ is the interleaving
function.

%${\mathcal{I}}(.)$ is the intra-frame interleaving permutation moving in each branch the
%symbol at time $t$ to the time $\sigma_t$. Hence, in each frame the sequence of
%interleaved super-symbol transmitted on the $n$ antennas is $\underline{c}_I=\left[{\bf
%C}^{(\sigma_1)}, \ldots,{\bf C}^{(\sigma_N)}\right]$.
%
%%%%%%%%%%%%%%%%
%where $\sigma=\sigma_1,\sigma_2,\ldots,\sigma_N$ is a permutation of the integers $1,
%\ldots, N$, the sum is over all permutations, and $\sgn(\sigma)$ denotes the sign of the
%permutation.
%%%%%%%%%%%%%%%%%

The channel model includes \ac{AWGN} and multiplicative flat fading,
with Rayleigh distributed amplitudes assumed constant over blocks of
$B$ consecutive transmitted space-time symbols and independent from
block to block \cite{MceSta:84, Chi:J98, MalLei:99}. Perfect channel
state information is assumed at the decoder.

The transmitted super-symbol at time $\sigma_t$ goes through the
channel described by the $(n \times m)$ channel matrix ${\bf
H}^{(\sigma_t)}=\left\{h_{i,s}^{(\sigma_t)}\right\}$ with ${i=1,\ldots,n;\
s=1,\ldots,m}$, where $h_{i,s}^{(\sigma_t)}$ is the channel gain
between transmitting antenna $i$ and receiving antenna $s$ at time
$\sigma_t$.

In the \ac{BFC} model these channel matrices do not change for $B$
consecutive transmissions, so that we actually have only $L=N/B$
possible distinct channel matrix instances per
codeword\footnote{For the sake of simplicity we assume that $N$
and $B$ are such that $L$ is an integer.}. By denoting with ${\bf
\mathcal{Z}}=\left\{{\bf Z}_1, \ldots , {\bf Z}_L\right\}$ the set
of $L$ channel instances, we have
\begin{equation}
{\bf H}^{(\sigma_t)}= {\bf Z}_l \qquad {\text{for}}\
\sigma_t=(l-1)B+1,...,l B \qquad , \qquad \ l=1, \ldots, L \, .
\end{equation}
%
%Here, ${\bf H}_1 \ldots {\bf H}_L$ are the $L$ fading matrices of the \ac{BFC},  which are
%constants in blocks of $B$ consecutive time instants, and independent from block to block
%The number of fading blocks per codeword is $L$, so that $%
%N=L\cdot B$.
When the fading block  length, $B$, is equal to one, we have the
ideally interleaved fading channel (i.e., independent fading
levels from symbol to symbol), while for $L=1$ we have the
quasi-static fading channel (fading level constant over a
codeword); by varying $L$ we can describe channels with different
correlation degrees \cite{MceSta:84, Chi:J98, MalLei:99}.

At the receiving side the sequence of received signal vectors is
 $\underline{r}_I=\left({\bf R}^{(\sigma_1)},
\ldots,{\bf R}^{(\sigma_N)}\right)$, and after de-interleaving we
have
$\underline{r}={{\mathcal{I}}^{-1}}(\underline{r}_I)=\left({\bf
R}^{(1)}, \ldots,{\bf R}^{(N)}\right)$, where the received vector
at time $t$ is ${\bf R}^{(t)}=\left[ r_{1}^{(t)}r_{2}^{(t)} \cdots
r_{m}^{(t)}\right] ^{T}$ with components
\begin{equation}
r_{s}^{(t)}=\sqrt{E_{s}}
\sum_{i=1}^{n}h_{i,s}^{(t)}c_{i}^{(t)}+\eta_{s}^{(t)}, \; \;\;\;\;
s=1,\ldots, m\,. \label{eq:rs}
\end{equation}
In this equation $r_{s}^{(t)}$ is the signal-space representation
of the signal received by antenna $s$ at time $t$, the noise terms
$\eta _{s}^{(t)}$ are \ac{i.i.d.} complex Gaussian \acp{r.v.},
with zero mean and variance $N_{0}/2$ per dimension, and the
\acp{r.v.} $h_{i,s}^{(t)}$ represent the de-interleaved complex
Gaussian fading coefficients. Since we assume spatially
uncorrelated channels, these are \ac{i.i.d.} with zero mean and
variance $1/2$ per dimension, and, consequently, $|h_{i,s}^{(t)}|$
are Rayleigh distributed r.v.s with unitary power. %The extension to Ricean fading is
%straightforward.
%
%The de-interleaved fading matrix sequence is then $\underline{\bf
%\alpha}=\left[{\bf \Omega}^{(1)}, \ldots,{\bf
%\Omega}^{(N)}\right]$, where  is the channel matrix at time $t$,
%is obtained as
%$\underline{\bf\alpha}_I={\mathcal{I}}^{-1}(\underline{\bf h})$,
%being $\underline{\bf h}=\left[{\bf H}^{(1)}, \ldots,{\bf
%H}^{(N)}\right]$ the fading matrix sequence, whose components are
%
The constellations are multiplied by a factor $\sqrt{E_{s}}$  in
order to have a transmitted energy per symbol equal to $E_s$,
which is also the average received symbol energy (per transmitting antenna) due
to the normalization adopted on fading gains.

%\note{NON SO SE TENERE QUESTA PARTE}
The total energy
transmitted per super-symbol is $E_{s_T}=n E_s$ %, and the signal-to-noise ratio per
%received antenna can be defined as $E_{s_T}/N_0$. On the other hand,
and the energy transmitted per information bit is $E_b=E_s/(h R)$
where $h$ is the
number of bits per modulation symbol and $R$ is the code-rate. % of the space-time encoder.
Thus, with ideal pulse shaping the spectral efficiency is $n h
R$\, $[bps/Hz]$.
%\note{END}

%The number of fading blocks per codeword is denoted by $L$, so that $%
%N=L\cdot B$. When the block  length, $B$, is equal to one, we have
%the common 'ideally interleaved' fading channel, whereas for $L=1$
%we have the 'quasi-static' fading channel. By varying $L$ we can
%describe channels with different correlation degrees \cite{mcit}.

For the discussion in the following sections it is worthwhile to
recall that, over a Rayleigh fading channel, the system
achieves a diversity $\mathcal{D}$ if the asymptotic error
probability is
%
%\begin{equation}
$P_{e}\approx {K}{\left( \frac{E_{s}}{N_{0}}\right)
^{-\mathcal{D}}}$
%\end{equation}
%
 where $K$ is a constant depending on the asymptotic coding gain
\cite{SchBenSte:B96,WintSalGit:94}. In other words, a system with
diversity $\mathcal{D}$ is described by a curve of error
probability with a slope approaching $10/\mathcal{D}$ [dB/decade]
for large \ac{SNR}.

%%%%%%%%%%%%%%%%%%%%%%%%%%%%%%%%%%%%%%%%%%%%%%%%%%%%%%%%%%%%%%%%%%%%%%%%%%%%%%%%
\section{A Pragmatic approach to space-time codes}
%%%%%%%%%%%%%%%%%%%%%%%%%%%%%%%%%%%%%%%%%%%%%%%%%%%%%%%%%%%%%%%%%%%%%%%%%%%%%%%%
\label{sec:pstc}

%\note{MARCO}

%\note{LA ANTICIPEREI QUI (VA RIFORMULATA)}

\begin{figure}
%\centerline{\includegraphics[width=0.95\linewidth,draft=false]
%    {schemePSTC.eps}}
{\tiny \centerline{\input{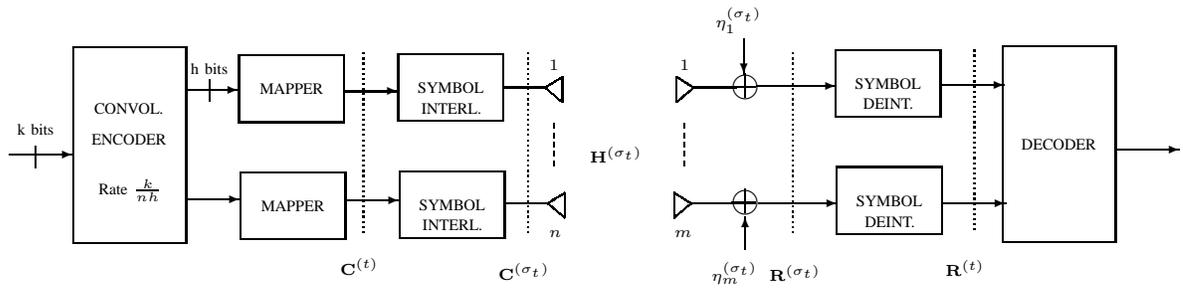}}} \vskip -8cm
\caption{Equivalent low-pass scheme for the proposed pragmatic
 space-time codes.} \label{fig:schemePSTC}
\end{figure}

In this section we present what we called
\acl{P-STC}, a low-complexity architecture for \ac{STC} %with
%\ac{BPSK} or \ac{QPSK} constellations,
that allows an easy code design and optimization over fading
channels \cite{ChiConTra:C01a}. The ''pragmatic'' approach %(the
%name intends to recall the similarity with what proposed in
%\cite{VitWolZehPad:89} for trellis-coded modulation)
consists in
using common convolutional codes as space-time codes, with the
architecture presented in Fig.~\ref{fig:schemePSTC}. Here, $k$
information bits are encoded by a convolutional encoder with rate
$k/(n h)$. The $n h$ output bits are divided into $n$ streams, one
for each transmitting antenna, of \ac{BPSK} ($h=1$) or \ac{QPSK}
($h=2$) symbols that are obtained from a natural (Gray) mapping of
$h$ bits. By natural mapping we mean that for \ac{BPSK} an
information bit $b\in\{0,1\}$ is mapped into the antipodal symbol
$c=2b-1$, giving $c\in\{-1,+1\}$; for \ac{QPSK} a pair of
information bits $a,b$ is mapped into a complex symbol
$c=(2a-1)/\sqrt{2}+j (2b-1)/\sqrt{2}$, giving $c\in\{\pm
1/\sqrt{2} \pm j /\sqrt{2}\}$, with $j=\sqrt{-1}$. Then, each stream of symbols is
eventually interleaved\footnote{In this paper we focus our
attention on symbol interleaving: bit interleaving is addressed in
\cite{ChiConTra:C02b}.}.
%
%As a result, for each group of $k$ information bits a super-symbol
%is transmitted, with an overall spectral efficiency $S_e=k$
%[bit/s/Hz].

We indicate the \ac{STC} obtained with this scheme as $(nh,k,\mu)$
$n$-\ac{P-STC}, where $\mu$ is the encoder constraint length and
the associated trellis has $N_{s}=2^{k(\mu-1)}$ states. For
example, we report in Fig.~\ref{fig:schemePSTCtxbpsk} the four
states (2,1,3) $2$-\ac{P-STC}  encoder scheme for $n=2$
transmitting antennas and \ac{BPSK} modulation, obtained with a
rate $1/2$ convolutional encoder with generator polynomials
$(5,7)_8$. We can describe \ac{P-STC} by using the trellis of the
encoder (the same as for the \ac{CC}), labelling the generic
branch from state $S_i$ to state $S_j$ with the super-symbol
$\widetilde{\bf C }_{S_i\rightarrow S_j
}=[\widetilde{c}_1,\ldots,\widetilde{c}_n]^T$, where for \ac{BPSK}
$\widetilde{c}_l$ is the output of the $l-th$ generator (in
antipodal version). In Fig.~\ref{fig:schemePSTCtxbpsktrellis} we
report the trellis for the \ac{P-STC} in
Fig.~\ref{fig:schemePSTCtxbpsk}.

\begin{figure}
{\small \centerline{\input{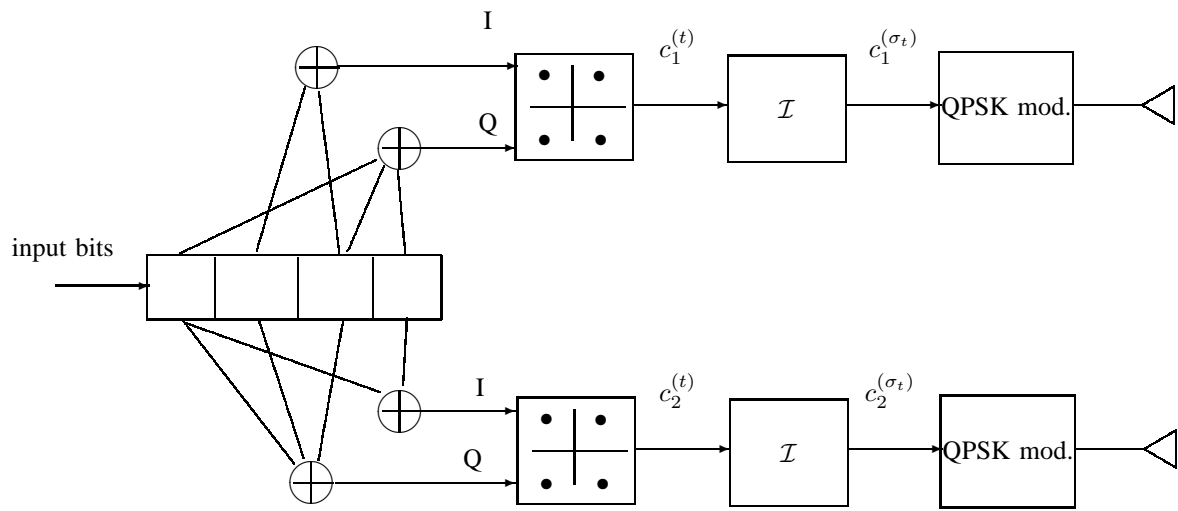}}} \vskip -16cm
\caption{Example of $2\ bps/Hz$ pragmatic space-time encoder for
$n=2$ transmitting antennas and QPSK modulation, obtained with a
rate $2/4$ convolutional encoder with $4$ states and generator
polynomials $(06,13,11,16)_8$.} \label{fig:schemePSTCtxqpsk}
\end{figure}

Similarly, in Fig.~\ref{fig:schemePSTCtxqpsk} we report the
 $4$ states (4,2,2)~2-\ac{P-STC} encoder scheme for $n=2$
transmitting antennas and \ac{QPSK} modulation, obtained with a
rate $2/4$ convolutional encoder with generator polynomials
$(06,13,11,16)_8$.

\begin{figure}
{\small \centerline{\input{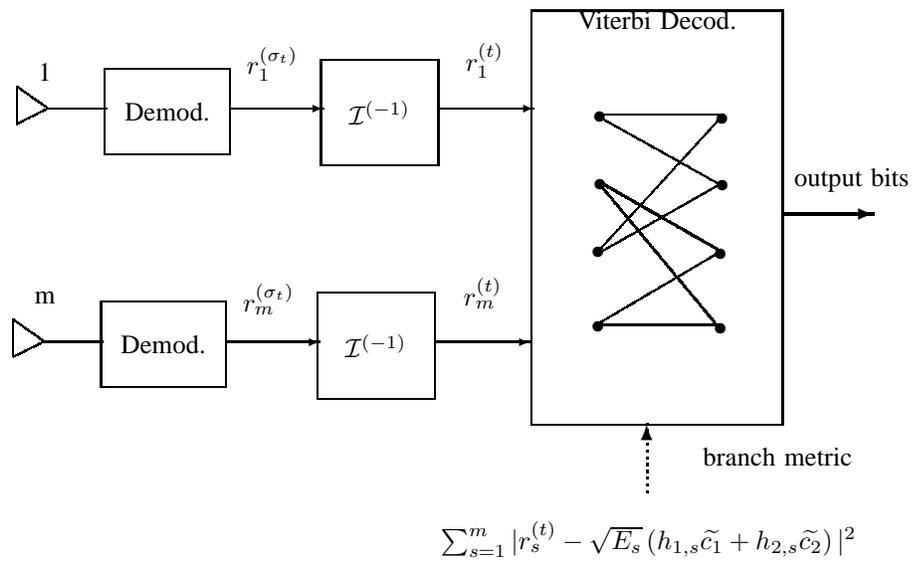}}} %\vskip -18cm
\caption{Receiver structure for the proposed pragmatic space-time
codes. In the figure the Viterbi decoder is the usual for the
convolutional code adopted in transmission, with the only change
that the metric on a generic branch is, for $n=2$, $\sum_{s=1}^m |
r_s^{(t)}-\sqrt{E_s} \left( h^{(t)}_{1,s} \widetilde{c_1} +
h^{(t)}_{2,s} \widetilde{c_2}\right) |^2$, being $\widetilde{c_1},
\widetilde{c_2}$ the two symbols associated to the branch.}
\label{fig:schemePSTCrx}
\end{figure}

It is clear now that with the pragmatic architecture the \ac{ML} decoder
is the usual Viterbi decoder for the convolutional encoder adopted
(same trellis), with a simple modification of the branch metrics.
For example, in Fig.~\ref{fig:schemePSTCrx} we show the receiver
architecture for the previous \ac{P-STC}, that simply consists in
the usual Viterbi decoder for the convolutional code adopted in
transmission, with the only change that the metric on a generic
trellis branch is $\sum_{s=1}^m | r_s^{(t)}-\sqrt{E_s} \left(
h^{(t)}_{1,s} \widetilde{c}_1 + h^{(t)}_{2,s}
\widetilde{c}_2\right) |^2$, being $\left\{\widetilde{c}_i\right\}$ the set of length $n$ of the output symbols labelling the branch. In general, for $n$
transmitting antennas, the branch metric for the Viterbi decoder
is
\begin{equation}\label{eq:metrics}
\sum_{s=1}^m | r_s^{(t)}-\sqrt{E_s} \sum_{i=1}^{n} h^{(t)}_{i,s}
\widetilde{c_i} |^2 \,.
\end{equation}
%
%In summary, t
Thus, we can resume the advantages of \ac{P-STC} with respect to
\ac{STC} as in the following: %in \cite{TarSesCal:98} are:

\begin{itemize}
\item  the encoder is a common convolutional encoder;

\item  the (Viterbi) decoder is the same as for a convolutional code, except
for a change in the metric evaluation;

\item  \ac{P-STC} are easy to study and optimize, even over \ac{BFC}.
\end{itemize}

These aspects will be further investigated in the next sections.

%While the first two properties give practical, implementation
%benefits, from a theoretical point of view the last property is
%also important. In fact, since these codes are wide-sense
%geometrically uniform (as will be shown later), the search for
%good generator polynomials is feasible even with long codewords
%and fast fading channels.
%
%(ARRIVATO FINO A QUI 30.5.06)
%
% This is possible by using the methodology previously
%presented to exhaustively investigate the whole class of possible
%codes, choosing those polynomial generators giving, for a given
%frame length, rate, channel memory $L$, and interleaving, the
%highest diversity gain and the best gain factor. For slow fading
%channels ($L=1$) since the union bound may be large, the approach
%of \cite{TarSesCal:98} can be used, looking, in the modified
%transfer function (\ref{T(D)}), at the terms with minimum product
%of eigenvalues instead of the gain factor in (\ref{CodingGain}).

%%%%%%%%%%%%%%%%%%%%%%%%%%%%%%%%%%%%%%%%%%%%%%%%%%%%%%%%%%%%%%%%%%%%%%%%%%%%%%%%
\section{ The pairwise error probability for space-time codes over \ac{BFC}}
%%%%%%%%%%%%%%%%%%%%%%%%%%%%%%%%%%%%%%%%%%%%%%%%%%%%%%%%%%%%%%%%%%%%%%%%%%%%%%%%
%\note{MARCO}

In this section we address the performance analysis for the general
class of \ac{STC} over \ac{BFC}.

Given the transmitted codeword $\underline{c}$, the PEP, that is  the probability that the \ac{ML} decoder
chooses the codeword $\underline{g} \neq \underline{c}$,
conditional to the set of fading levels ${\bf \mathcal{Z}}$, can
be written as

\begin{equation}
\PX{\underline{c}\rightarrow \underline{g}  | {\bf \mathcal{Z}} } =\frac{1}{2}\ \erfc\sqrt{\frac{E_{s}}{4N_{0}}%
d^{2}\left( \underline{c},\underline{g} | {\bf \mathcal{Z}}
\right) } \,, \label{peH}
\end{equation}
%
%Remember that $\alpha _{i,s}^{(u)}$ is the fading coefficient from the $i^{th} $ transmitting antenna to the $s^{th}$ receiving antenna, at time $u.$
where $ \erfc(x)\triangleq \frac{2}{\sqrt{\pi}}\int_x^\infty
e^{-t^2} dt$ is the complementary Gaussian error function, and the
conditional Euclidean squared distance at the channel output,
$d^{2}\left( \underline{c},\underline{g} | {\bf \mathcal{Z}}
\right)$, is given by \cite{TarSesCal:98}

%The distance in (\ref{peH}) is, following \cite{TarSesCal:98}:

\begin{equation}
d^{2}\left( \underline{c},\underline{g}|{\bf \mathcal{Z}} \right)
=\sum_{t=1}^{N} \sum_{s=1}^{m}\left|
\sum_{i=1}^{n}h_{i,s}^{(t)}\cdot \left(
c_{i}^{(t)}-g_{i}^{(t)}\right) \right| ^{2}\,. \label{eq:d2cgZ}
\end{equation}
To specialize this expression to the \ac{BFC} we first rewrite the
squared distance as follows
\begin{eqnarray}
d^{2}\left( \underline{c},\underline{g}|{\bf \mathcal{Z}} \right) &=&\sum_{t=1}^{N}
\sum_{s=1}^{m}{\bf h }_{s}^{(t)}\left( {\bf C}^{(t)}-{\bf G}^{(t)}\right)
\cdot  \left( {\bf C}^{(t)}-{\bf G}^{(t)}\right) ^{H}{\bf h }_{s}^{(t)H\,} \nonumber\\
&=&\sum_{t=1}^{N} \sum_{s=1}^{m}{\bf h }_{s}^{(t)}{%
{\bf A}}^{(t)}(\underline{c},\underline{g}){\bf h}_{s}^{(t)H\,} \,, \label{d2Omega}
\end{eqnarray}
where ${ \bf h }_{s}^{(t)}=\left[
h_{1,s}^{(t)},h_{2,s}^{(t)},...,h_{n,s}^{(t)}\right] $ is the
$(1\times n)$ vector of the fading coefficients related to the
receiving antenna $s$, and ${\bf C}^{(t)},{\bf G}^{(t)}$ are the
super-symbols at time $t$ in the sequence $\underline{c},$ and
$\underline{g}$, respectively.
In \eqref{d2Omega} the ($n\times n$) matrix
$${{\bf A}}^{(t)}(\underline{c},\underline{g})=\left( {\bf C}^{(t)}-%
{\bf G}^{(t)}\right) \left( {\bf C}^{(t)}-{\bf G}%
^{(t)}\right) ^{H}$$
with elements
$$A^{(t)}_{p,q}=\left( c_{p}^{(t)}-g_{p}^{(t)}\right) \left(
c_{q}^{(t)}-g_{q}^{(t)}\right) ^{*}$$
is Hermitian and non-negative definite\footnote{This can be simply
verified by noting that, since ${\bf A}$ can be written as  ${{\bf
A}}={\bf y  y}^{H}$, for every $(1\times n)$ vector ${\bf x}$ we
have ${\bf x A x}^{H}={\bf x y  y^{H} x^{H}=|| x y}||^{2}\geq
0$.}.

Due to the \ac{BFC} assumption, for each frame and each receiving
antenna the fading channel is described by only $L$ different
vectors
%${\bf h }_{s}^{(t)}$, that we indicate as $%
%{\bf z}_{s}^{(1)},{\bf z}_{s}^{(2)},..,{\bf z}_{s}^{(L)},$
%i.e.
${ \bf h }_{s}^{(t)} \in \left\{ {\bf z}_{s}^{(1)},%
{\bf z}_{s}^{(2)}, \ldots ,{\bf z}_{s}^{(L)}\right\} $ , $s=1,
\ldots ,m$, where ${\bf z}_{s}^{(l)}$ is the $s$-th row of ${\bf
Z}_l$. By grouping these vectors, we can rewrite \eqref{d2Omega}
as

\begin{equation}
d^{2}\left( \underline{c},\underline{g} | {\bf \mathcal{Z}}
\right) =\sum_{l=1}^{L}\sum_{s=1}^{m}{\bf z}_{s}^{(l)}{%
{\bf F}}^{(l)}(\underline{c},\underline{g}){\bf z}_{s}^{(l)\,H} \,, \label{d2F}
\end{equation}
where
\begin{equation}\label{eq:Fcg}
{{\bf F}}^{(l)}(\underline{c},\underline{g}) \triangleq \sum_{t\in T(l)}{{\bf A}}%
^{(t)}(\underline{c},\underline{g})=\sum_{t\in T(l)}\left( {\bf
C}^{(t)}-{\bf G}^{(t)}\right) \cdot \left( {\bf C}^{(t)}-{\bf
G}^{(t)}\right) ^{H} \,\,\, l=1,\ldots,L
\end{equation}
and $T(l)\triangleq \{t: {\bf H}^{(\sigma_t)}= {\bf Z}_l \}$ %=\left\{t:{\bf H}^{(t)}={\bf H}_l\right\}$
is the set of indexes $t$ where the channel fading gain matrix is
equal to ${\bf Z}_{l}$. This set depends on the interleaving
strategy adopted. Note that in our scheme
(Fig.~\ref{fig:schemePSTC}) the interleaving is done
``horizontally'' for each transmitting antenna and that the set
$T(l)$ is independent on $s$, that means, in other words, that the
interleaving rule is the same for all antennas.

The matrix ${{\bf F}}^{(l)}(\underline{c},\underline{g})$ is also Hermitian non-negative definite, being the sum of
Hermitian non-negative definite matrices.
It has, therefore, real non-negative eigenvalues.
Moreover, it can be written as ${{\bf F}}^{(l)}(\underline{c},\underline{g})={{\bf U}}^{(l)}{%
{\bf \Lambda }}^{(l)}{{\bf U}}^{(l)H}$, where $%
{{\bf U}}^{(l)}$ is a unitary matrix and ${%
{\bf \Lambda }}^{(l)}$ is a real diagonal matrix, whose diagonal
elements $\lambda _{i}^{(l)}$ with $i=1,\ldots,n$ are the eigenvalues of ${%
{\bf F}}^{(l)}(\underline{c},\underline{g})$ counting multiplicity. Note that ${%
{\bf F}}^{(l)}$ and its eigenvalues $\lambda _{i}^{(l)}$ are a
function of $\underline{c}-\underline{g}$. As a result, we can
express the squared distance $d^{2}\left(\underline{c},\underline{g}| {\bf \mathcal{Z}} \right)$ by
utilizing the eigenvalues of ${{\bf
F}}^{(l)}(\underline{c},\underline{g})$ as follows:

\begin{eqnarray}\label{eq:d2eigen}
d^{2}\left( \underline{c},\underline{g}|{\bf \mathcal{Z}}
\right) &=&\sum_{l=1}^{L}\sum_{s=1}^{m}{\bf z}_{s}^{(l)}{%
{\bf U}}^{(l)}{{\bf \Lambda }}^{(l)}{%
{\bf U}}^{(l)H}{\bf z}_{s}^{(l)\,H}  \nonumber \\
&=&\sum_{l=1}^{L}\sum_{s=1}^{m}{\bf B}_{s}^{(l)}{{%
\bf \Lambda }}^{(l)}{\bf B}_{s}^{(l)\,H}  \nonumber \\
&=&\sum_{l=1}^{L}\sum_{s=1}^{m}\sum_{i=1}^{n}\lambda
_{i}^{(l)}\left| \beta _{i,s}^{(l)}\right| ^{2}
\end{eqnarray}
where ${\bf B}_{s}^{(l)}=\left[ \beta _{1,s}^{(l)},\beta
_{2,s}^{(l)},...,\beta _{n,s}^{(l)}\right] ={\bf z}_{s}^{(l)}%
{{\bf U}}^{(l)}$.

The difference between \eqref{eq:d2eigen} and the similar
expression reported in \cite{TarSesCal:98} is that, through \eqref{eq:Fcg}, the eigenvalues
in \eqref{eq:d2eigen} are referred to the portions of the coded
sequences with a given fading level. %So,
%these eigenvalues are $L$ instead of $N$: for the particular case
%of ideally interleaved channel, $L=N$, equations
%\eqref{eq:Fcg}\eqref{eq:d2eigen} result in those in
%\cite{TarSesCal:98}.

Since ${{\bf U}}^{(l)}$ represents a unitary transformation, ${\bf
B}_{s}^{(l)}$ has the same statistical description of ${\bf
z}_{s}^{(l)}$. Hence, in the case of Rayleigh distribution, ${\bf
B}_{s}^{(l)}$ has independent, complex Gaussian elements, with
zero mean and variance 1/2 per
dimension. Moreover, for \ac{BFC}, vectors ${\bf B}_{s}^{(l)}$ and ${\bf B}%
_{s}^{(j)} $ are independent $\forall l\neq j$.
%By introducing the position $\Gamma =E_{s}/4N_{0}$, the
Hence, the unconditional \ac{PEP} becomes

% \begin{equation}
% P\left( \underline{c}\rightarrow \underline{g}|\left\{ \alpha
% _{i,s}^{(t)}\right\} \right) =\frac{1}{2}erfc\sqrt{\Gamma
% \sum_{s=1}^{m}\sum_{v=1}^{L}\sum_{i=1}^{n}\lambda _{i}^{(v)}\left| \beta
% _{i,s}^{(v)}\right| ^{2}}  \label{PeBetai,s}
% \end{equation}

% and, averaging over the fading,

\begin{equation}
\PX{ \underline{c}\rightarrow \underline{g} } =\EX{ \frac{1}{2}%
\erfc\sqrt{\frac{E_s}{4 N_0}
\sum_{s=1}^{m}\sum_{l=1}^{L}\sum_{i=1}^{n}\lambda _{i}^{(l)}\left|
\beta _{i,s}^{(l)}\right| ^{2}}} \label{PairEerfc}
\end{equation}
where $\EX{.}$ indicates expectation with respect to fading. By
evaluating the asymptotic behavior for large \ac{SNR} of
\eqref{PairEerfc} we obtain (see \cite{ChiConTra:04})
\begin{equation}
P\left( \underline{c}\rightarrow \underline{g}\right) \leq
K(m\eta)\left[\prod_{l=1}^{L}\prod_{i=1}^{^{\eta _{l}}}\lambda
_{i}^{(l)}\left(\frac{E_s}{4 N_0}\right)^{\eta }\right] ^{-m}
\label{Pair}
\end{equation}
where\footnote{A looser bound can be obtained by observing that
$K({d}) \leq 1/4$.} $$K({d})=\frac{1}{2^{2 {d}}} \binom{2{ d
}-1}{{ d }} \,,$$ the integer
$\eta_{l}=\eta_{l}(\underline{c},\underline{g})$ is the number of
non-zero eigenvalues of ${{\bf
F}}^{(l)}(\underline{c},\underline{g})$, and $\eta$ (that we can
call the pairwise transmit diversity) is the sum of the ranks of ${{\bf
F}}^{(l)}(\underline{c},\underline{g})$, i.e.
\begin{equation}
\eta =\eta (\underline{c},\underline{g})=\sum_{l=1}^{L}\rank%
\left[ {{\bf F}}^{(l)}(\underline{c},\underline{g})\right]
=\sum_{l=1}^{L} \eta _{l}\,.
\end{equation}
The \ac{PEP} between $\underline{c}$ and $\underline{g}$ shows a
diversity $m \eta$ that is the product of transmit and receive diversity.

Equation (\ref{Pair}) can be seen as the generalization to
\ac{BFC} of the \ac{PEP} for the quasi-static channel in
\cite{TarSesCal:98}: for the \ac{BFC}, to obtain the \ac{PEP} we
must compute the product and the number of non-zero eigenvalues of
the set of suitably defined matrices ${{\bf
F}}^{(l)}(\underline{c},\underline{g}) \,$, $l=1,\ldots,L$,
accounting, through \eqref{eq:Fcg}, for the number of fading
levels per codeword and for the
interleaving rule. % for each possible pair
%$\underline{c},\underline{g}$.
The analysis is valid for \ac{STC} and will be applied also to \ac{P-STC}.

%%%%%%%%%%%%%%%%%%%%%%%%%%%%%%%%%%%%%%%%%%%%%%%%%%%%%%%%%%%%%%%%%%%%%%%%%%%%%%%%
\section{Error probability analysis for \ac{STC} over \ac{BFC}}
%%%%%%%%%%%%%%%%%%%%%%%%%%%%%%%%%%%%%%%%%%%%%%%%%%%%%%%%%%%%%%%%%%%%%%%%%%%%%%%%
%\note{MARCO}

% For geometrically uniform codes, that have all the
%distance profiles from any codeword to all other codewords equal,
%the performance can be evaluated by assuming the transmission of a
%particular codeword $\underline{z}$, that can be the 'all-zero'
%codeword generated when all the information bits are $0$. This is
%essentially due to the fact that in flat fading channels the error
%probability as in (\ref{Pair}) depends directly on the difference
%between codewords, e.g. $\underline{c}-\underline{g}$.
%
%Nota: the code at the output of the multiple input multiple output
%channel may lose the geometrical uniformity; however, from the
%performance point of view, is important the uniformity of the code
%before the channel.

Given the transmitted codeword $\underline{c}$ the frame error
probability, $P_{w}(\underline{c})$, can be bounded through the
union bound as
\begin{equation}
P_{w}(\underline{c})\leq \sum_{\underline{g}\neq \underline{c}}
\PX{\underline{c}\rightarrow \underline{g}}\,,
\label{pcod}
\end{equation}
that, using \eqref{Pair}, gives for large SNR
%By introducing the error sequence
%$\underline{e}=\underline{g}-\underline{z}$, the unconditional
%frame error probability can be bounded for large \ac{SNR} as
%
\begin{equation}
P_{w}(\underline{c})\le \sum_{\underline{g}\neq
\underline{c}} K(m \eta) \left[\prod_{l=1}^{L}\prod_{i=1}^{{\eta
_{l}}}\lambda _{i}^{(l)}\left(\frac{E_s}{4 N_0}\right)^{\eta
}\right] ^{-m} \,\,\, , \label{CEP1}
\end{equation}
where the dominant terms are those with minimum $\eta$. It should
be reminded that the parameters  ${\eta _{l}}$ and $\lambda
_{i}^{(l)}$ depend on codewords $\underline{c}$ and
$\underline{g}$. By retaining dominant terms only, the conditional
asymptotic error probability bound becomes
\begin{equation}
P_{w_\infty }(\underline{c}) \approx K(m \eta _{\min }(\underline{c})) \left(\frac{E_s}{4 N_0}\right)^{-\eta _{\min }(\underline{c})\cdot m}\sum_{%
\underline{g}\in {\mathcal{E}}(\underline{c},\eta _{\min }(\underline{c}))}\left[\prod_{l=1}^{L}\prod_{i=1}^{{\eta _{l}}}\lambda
_{i}^{(l)}\right] ^{-m} \label{CEP2}
\end{equation}
where $\eta _{\min }(\underline{c})=\min_{\underline{g}}\eta
(\underline{c},\underline{g})$ and ${\mathcal{E}}(\underline{c},x)=\left\{
\underline{g}\neq \underline{c}:\eta
(\underline{c},\underline{g})=x\right\} $ is the set of
codeword sequences at minimum diversity. The asymptotic bound
shows that the achievable diversity (also called diversity gain), $\eta
_{\min}(\underline{c})\cdot m $, increases linearly with the number of receiving
antenna.
 Note that, here, the transmit diversity order
$\eta_{\min}(\underline{c})$ has the same significant role of the
code free distance, $d_f$, in \ac{AWGN} channels.

When dealing with codes for which the conditional error
probability, $P_w(\underline{c})$, does not depend on the
transmitted codeword $\underline{c}$ (see also the discussion in a
following subsection), the unconditional error probability can be
evaluated by arbitrarily selecting a reference codeword $\underline{c}_0$.
In the same way we may use $P_w(\underline{c}_0)$  as a bound for
those codes for which we can prove that $\underline{c}_0$ is the
worst case reference codeword. However, in general the error
probability bound must be evaluated as
%\tilde{}
\begin{equation}
\tilde{P}_{w}=\sum_{\underline{c}} \PX{\underline{c}}
P_w(\underline{c}) \leq
\sum_{\underline{c}}\sum_{\underline{g}\neq \underline{c}}
\PX{\underline{c}} \PX{\underline{c}\rightarrow \underline{g}
}\,, \label{pcodgen}
\end{equation}
where $\PX{\underline{c}}$ is the probability of transmitting the
codeword $\underline{c}$ (i.e., for \ac{P-STC}, equal to $2^{-kN}$ for
equiprobable input bit sequence and  $2^{-k(N-\mu+1)}$ for a zero
tailed code). By using \eqref{CEP2}, and by observing that the retained dominant terms are those with transmit diversity  $\tilde{\eta} _{\min }=\min_{\underline{c}}\eta _{\min
}(\underline{c})$, the asymptotic error
probability bound can be written
\begin{equation}
\tilde{P}_{w_\infty } \approx K(\tilde{\eta}_{\min} m)
\left(\frac{E_s}{4 N_0}\right)^{-\tilde{\eta} _{\min}\cdot m}
\sum_{\underline{c}} \PX{\underline{c}} \sum_{\underline{g}\in
{\mathcal{E}}(\underline{c},\tilde{\eta} _{\min })}
\left[\prod_{l=1}^{L}\prod_{i=1}^{{\eta
_{l}}}\lambda _{i}^{(l)}\right] ^{-m} \,. \label{CEP3}
\end{equation}
From \eqref{CEP3} we observe that the
asymptotic performance of \ac{STC} over \ac{BFC}
%given the transmitted codeword $\underline{c}$
depends on both the achievable diversity, $\tilde{\eta} _{\min }\cdot
m$, and the performance factor
\begin{equation}
\tilde{F}_{\min}(m)=\sum_{\underline{c}} \PX{\underline{c}}
F_{\min}(\underline{c},m)
\triangleq \sum_{\underline{c}}\PX{\underline{c}}\sum_{%
\underline{g}\in {\mathcal{E}}
(\underline{c},\tilde{\eta} _{\min })}
\left[\prod_{l=1}^{L}\prod_{i=1}^{{\eta _{l}}}\lambda
_{i}^{(l)}\right] ^{-m}\,, \label{CodingGain}
\end{equation}
which is related to the coding gain in \eqref{CEP3}.

Note also that $\tilde{\eta} _{\min }$ and the weights
$\prod_{l=1}^{L}\prod_{i=1}^{^{\eta _{l}}}\lambda _{i}^{(l)}$ for each $\underline{c}$ and  $\underline{g}$ do
not depend on the number of receiving antennas. Therefore, when a
code is found to reach the maximum diversity $\tilde{\eta} _{\min
}$ in a system with one receiving antenna, the same code reaches
the maximum diversity $\tilde{\eta} _{\min }\cdot m$ when used
with multiple receiving antennas. However, due to the presence of
the exponent $m$ in  each term of the sum in \eqref{CodingGain},
the best code (i.e., the code having the smallest performance
factor) for a given number of antennas is not necessarily the best
for a different number of receiving antennas. Thus, a search for
optimum codes in terms of both diversity and performance factor
must in principle be pursued for each $m$.

To summarize, the derivation of the asymptotic behavior of a
given \ac{STC} with a given length requires computing the %, for each
%non zero error sequence,
 matrices ${{\bf F}}^{(l)}(\underline{c},\underline{g})$ in \eqref{eq:Fcg} with
their rank and product of non-zero eigenvalues. In relation with
\cite{CaiVit:98}, we also observe that:
\begin{itemize}
\item By restricting in the bound the set of sequences $\underline{g}$ to those corresponding to paths in
the trellis diagram of the code diverging only once from the path
of codeword $\underline{c}$, the union bound becomes tighter.
\item By restricting in the bound the set of sequences $\underline{g}$ to those corresponding to paths in
the trellis diagram of the code diverging only once and only at
time $t$ from the path of codeword $\underline{c}$, we obtain the
first event error probability at time $t$. In the particular case
of periodical interleaving over the \ac{BFC} we can use the first
event error probability at $t=0$ to obtain a simpler but
looser union bound, in the following form:\footnote{This is also
known as first event error probability analysis.}
\begin{equation}
\tilde{P}_w \leq N \sum_{\underline{c}} \PX{\underline{c}} \sum_{\underline{g}\in
{\mathcal{E}_0} (\underline{c})}
\PX{\underline{c}\rightarrow \underline{g} }\,,
\label{pe_fe}
\end{equation}
where $ {\mathcal{E}_0}(\underline{c})$ is the set of codewords $\underline{g}$ restricted to the first event error; this set must be used to evaluate the
asymptotic performance \eqref{CEP2} and the performance factor in
\eqref{CodingGain}.
\item From the error probability bound we can easily obtain an approximation by truncating the number
of terms in the asymptotic expression \eqref{CEP3} to the most
significant terms, i.e, by keeping those terms with product of the
non-zero eigenvalues smaller than a selected threshold $\delta_P$,
or those terms corresponding to pairs
$(\underline{c},\underline{g})$ with Hamming distance smaller than
a selected threshold $\delta_H$.
\end{itemize}

%%%%%%%%%%%%%%%%%%%%%%%%%%%%%%%%%%%%%%%%%%%%%%%%%%%%%%%%%%%%%%%%%%%%%%%%%%%%%%%%
\subsection{The New Concept of Space-Time Generalized Transfer Function for \ac{P-STC} in \ac{BFC}}
\label{sec:STGTF}
%%%%%%%%%%%%%%%%%%%%%%%%%%%%%%%%%%%%%%%%%%%%%%%%%%%%%%%%%%%%%%%%%%%%%%%%%%%%%%%%
%\note{VELIO}

The evaluation of the error probability bound for \ac{P-STC} can be carried out
in an effective way, by extending the methodology in \cite{ChiConTra:04} for \ac{CC} over \ac{BFC}. This leads to the definition of the novel concept of \ac{ST-GTF} for
\ac{BFC}.  With respect to \ac{CC} some modifications are required,
as explained here, to account for the space-time fading channel.

 In order to define the \ac{ST-GTF} let us
first introduce the error sequences and discuss their role in the
evaluation of error probability. \ac{P-STC} are built using common
binary convolutional codes, therefore they are group-trellis codes
\cite{Sch:91}. If we consider the input bit sequences
$\underline{b}_c$ and $\underline{b}_g$, of length $kN$, that
generate the output codewords $\underline{c}$ and $\underline{g}$,
and define\footnote{With $\oplus$ we denote the element-wise
binary sum.} $\underline{e}=\underline{b}_c\oplus\underline{b}_g$
as the input error sequence
for the transmitted codeword $\underline{c}$ and decoded codeword $\underline{g}$, we can say that:\\
- by encoding the input bit sequence $\underline{e}$ with the \ac{P-STC} encoder we obtain a valid codeword; \\
- given a transmitted sequence $\underline{c}$ (or the
corresponding input $\underline{b}_c$), the whole set of error
sequences can be represented with
the same trellis diagram used to describe the code. The all-zero path in this case describes the event of correct decoding.\\
Having this in mind, we can rewrite the frame error probability
bound as
\begin{equation}
\tilde{P}_{w} \leq \sum_{\underline{e}\neq
\underline{0}}\sum_{\underline{b}_c} \PX{\underline{b}_c}
\PX{{\cal C}
(\underline{b}_c) \rightarrow {\cal C}
(\underline{b}_c\oplus\underline{e}) }\,, \label{pcod_e}
\end{equation}
where ${\cal C}(.)$ is the encoding function, $\PX{\underline{b}_c}$ is the probability to encode the input bit sequence ${\underline{b}_c}$, and
$\PX{{\cal C}
(\underline{b}_c) \rightarrow {\cal C}
(\underline{b}_c\oplus\underline{e}) }$ is the \ac{PEP}
related to input sequence $\underline{b}_c$ and input error
sequence $\underline{e}$. As before, the bound is preserved by
restricting the set of error sequences to those  represented by
paths in the trellis diverging only once from the all-zero path.
The bound is also simplified by considering the error paths
diverging from the all-zero path at $t=0$.

Thus, within this framework we can be proceed with the following
steps to the definition and the exploitation of the \ac{ST-GTF} of
the code, for which an example is given in Appendix:
%(conviene introdurre e seguire un esempio di codice)

%\begin{itemize}
a) construction of the error trellis diagram of the \ac{P-STC},
starting from the trellis diagram of length $N$ branches
describing the \ac{P-STC}. As observed, this trellis can be used
both for the set of input sequences $\underline{b}_c$ and for the
set of error sequences $\underline{e}$ (they both have the same
trellis diagram of the convolutional code but with different
meanings of input and output sequences). Let us denote with
$s_b^{(t)}$ and $s_e^{(t)}$ the binary vectors
representing the generic state at time $t$ when the trellis is
referred to the input sequence and to the error sequence,
respectively. Recall that in our notation the number of states is
$N_{s}=2^{k(\mu-1)}$, where $\mu$ is the constraint length of
the convolutional code. Let us build the error trellis diagram
according to \cite[Chap. 12]{BenBig:B98} by labeling the edges of
the trellis referred to the error sequences $\underline{e}$ with $
N_{s}\times N_{s}$ matrices $ {\bf
E}(s_e^{(t)},s_e^{(t+1)})$, whose generic element $i,j$
depends on the label $\widetilde{{\bf C}}_{S_i\rightarrow S_j} $ of
the transition from state $S_i$  to state $S_j$, and is given by
$$
\widetilde{{\bf C}}_{S_i\rightarrow S_j}-\widetilde{{\bf
C}}_{s_e^{(t)}\oplus S_i\rightarrow s_e^{(t+1)} \oplus
S_j}
$$
where $\widetilde{{\bf C}}_{s_e^{(t)}\oplus S_i\rightarrow
s_e^{(t+1)} \oplus S_j}$ is the label of the transition from
state
$s_e^{(t)}\oplus S_i$ to $s_e^{(t+1)} \oplus S_j$.\\

%Let us build a product trellis by using the set of superstates $\Sigma_{be}^{(t)}=(\sigma_b^{(t)},\sigma_e^{(t)})$,
% by connecting a superstate $\Sigma_{be}^{(t)}$  with a superstate $\Sigma_{be}^{(t+1)}$  if both the transitions
%$\sigma_b^{(t)}\longrightarrow \sigma_b^{(t+1)} $ $\sigma_e^{(t)}\longrightarrow \sigma_e^{(t+1)} $ exist in in the original trellis,
%and  by labeling this edge with the difference $ {\bf C}^{(t)}-{\bf G}^{(t)}$ where $ {\bf C}^{(t)}$ is the output supersymbols corresponding
%to the transition $\sigma_b^{(t)}\longrightarrow \sigma_b^{(t+1)} $ and ${\bf G}^{(t)}$ is the output supersymbols corresponding
%to the transition $\sigma_b^{(t)}\oplus \sigma_e^{(t)}\longrightarrow \sigma_b^{(t+1)} \oplus\sigma_e^{(t+1)}$

b) Construction of a modified error trellis diagram by labelling
the generic transition $s_{e}^{(t)}
\rightarrow s_{e}^{(t+1)}$ of the error trellis with a new
matrix label $ {\bf E}'(s_e^{(t)},s_e^{(t+1)})$ whose
generic element $i,j$ is given by\footnote{In case of terminated
codes by means of zero tailing the term $2^{-k}$ has to be removed
for $t=N-\mu+1, ..., N-1$.}
\begin{equation}
\Delta_{t}^{2^{-k}{\bf A} ^{(t)}_{i\rightarrow j}}\,,
\end{equation}
where $\Delta_{1},\ldots,\Delta_{N}$ are indeterminates and ${\bf A}
^{(t)}_{i\rightarrow j}$ is the $n\times n$ matrix given by
$${\bf A}^{(t)}_{i\rightarrow j}=\left( \widetilde{{\bf C}}_{S_i\rightarrow S_j}-\widetilde{{\bf
C}}_{s_e^{(t)}\oplus S_i\rightarrow s_e^{(t+1)} \oplus
S_j}\right) \left( \widetilde{{\bf C}}_{S_i\rightarrow
S_j}-\widetilde{{\bf C}}_{s_e^{(t)}\oplus S_i\rightarrow
s_e^{(t+1)} \oplus S_j} \right) ^{H}\,.$$ This trellis
diagram, named {\it error trellis with error matrices}, depends on
the sequence of dummy variables ${\mathbf{\Delta}}=\left(\Delta_1,
\ldots,\Delta_{N}\right)$ related to the multiple-input fading
channel level as seen by each super symbol of the codeword. As
before, due to finite interleaving, for each realization there are
only a few different fading levels per frame. As an example, in
case of quasi-static channels only one indeterminate must be used.
In the opposite case of perfect symbol interleaving, although the
number of indeterminates could be taken equal to the frame length,
$N$, for the description of the average error probability over
fading only one indeterminate may be used.\\

c) Construction of the error trellis with error matrices for the
\ac{BFC} by using the same indeterminate  variable for super-symbols
subjected to the same fading gain. This can be simply done with the
position:
\begin{equation}
{\mathbf{\Delta}}={\mathcal{I}}^{-1}\left(\underbrace{D_1,\ldots,D_1}_{B
\, times}, \ldots, \underbrace{D_L, \ldots,D_L}_{B \,
times}\right)\,, \label{inter}
\end{equation}
which makes the trellis labels a function of the new set of dummy
variables ${\bf D}=\{D_1,...,D_L\}$, each related to
one of the $L$ fading levels.\\

d) Evaluation, by using standard techniques, of the transfer
function for the error trellis diagram as in \cite{ChiConTra:J04},
but with error matrices and by using the rules:
\begin{equation}
D_{l}^{{\bf{A}}_{1}}\cdot D_{l}^{{\bf{A}}%
_{2}}=D_{l}^{{\bf{A}}_{1}+{\bf{A}}_{2}}
\end{equation}
\begin{equation}
a\cdot D_{l}^{{\bf{A}}_{1}}+b\cdot D_{l}^{{%
\bf{A}}_{1}}=(a+b)D_{l}^{{\bf{A}}_{1}}
\end{equation}
\begin{equation}
a\cdot D_{l}^{{\bf{O}}}=1
\end{equation}
where ${\bf{A}}_{1}, {\bf{A}}_{2}$ are generic non-negative definite
matrices, ${\bf{O}}$ is the zero matrix, $a$ and $b$ are scalars.
 In fact, we have to define for each node of the error trellis with error matrices, that
is for each state $s_e^{(t)}=s$ at time $t$, a $N_s \times 1$ weighting vector polynomial
${\bf Q}_{s}({\bf D})$  which can be evaluated as the sum over all the
transitions reaching $s$ of the polynomials obtained by
multiplying each transition label, which is a $N_s \times N_s$ matrix, by the weight of the node at
time $t-1$ from which the branch departs. Next, if we set to $1$
the weight of the initial state of the trellis, denoted by $O_0$
(the zero-state at the time 0), we can obtain what we call the
\acf{ST-GTF} as
\begin{equation}
 \label{eq:TMD}
T_M({\bf D})=  {\bf Q}_{O_N}({\bf D})^T {\bf U}_0-1\,,
\end{equation}
where ${\bf U}_0^T =[1\; 0 \; 0 \dots 0]$, $O_N$ is the final state of the trellis (the zero-state at time $N$) and the contribution of the
correct sequence (the polynomial $1$) is subtracted.

In \eqref{eq:TMD} the \ac{ST-GTF} has the form of a polynomial in
the indeterminates $D_{1},\ldots,D_{L}$ with matrix exponents
%

%\lefteqn{T_M(D_{1},...,D_{L})= }
\begin{equation}
T_M(D_{1},...,D_{L})=\sum_{{\scriptsize ({\bf {F}}^{(1)},\ldots,{\bf {F}}^{(L)}) \neq ({\bf {0}},\ldots,{\bf {0}})
}}w({\bf {F}}%
^{(1)},\ldots,{\bf {F}}%
^{(L)}) \cdot D_{1}^{{\bf {F}}^{(1)}}\cdot \cdot \cdot D_{L}^{%
{\bf {F}}^{(L)}}  \label{T(D)}
\end{equation}
where each pairwise error event is characterized by a set of $L$
matrices $({\bf {F}}^{(1)},\dots,{\bf {F}}^{(L)}) \neq
({\bf {0}},\ldots,{\bf {0}})$, and $w({\bf {F}}^{(1)},\ldots,{\bf {F}}^{(L)})$ enumerates (including the weight
$\PX{\underline{b}_c}$) the error sequences producing ${\bf
{F}}^{(1)},\dots,{\bf {F}}^{(L)}$.
Among the terms in
\eqref{T(D)}, the most important are those related to
matrices ${\bf {F}}^{(1)},\ldots,{\bf {F}}^{(L)}$ having minimum diversity, that is, the
minimum value of $\sum_{l=1}^{L}\rank\left[ {\bf {F}}^{(l)}\right]$.
For these, it is important to evaluate the weight $\prod_{l=1}^{L}%
\prod_{i=1}^{^{\eta _{l}}}\lambda _{i}^{(l)}$ given by the product of the all
non-zero eigenvalues of ${\bf {F}}^{(1)},\ldots,{\bf {F}}^{(L)}$.\\

e) Symbolic substitution of the powers in the \ac{ST-GTF} with
distances, by using the linear operator defined as
\begin{equation}
 {\cal T} \left[\alpha \cdot \prod_{l=1}^{L} D_{l}^{{\bf {F}}^{(l)}} \right] =
 \alpha \cdot d_{1,1}^{\lambda_{1}^{(1)}}\cdots
 d_{1,\eta_1}^{\lambda_{\eta_1}^{(1)}}\cdots d_{L,1}^{\lambda_{1}^{(L)}}\cdots
 d_{L,\eta_L}^{\lambda_{\eta_L}^{(L)}}\,,
\end{equation}
where $\alpha\in\mathbb{R}$ is an arbitrary number
and $\lambda_{1}^{(l)}, \ldots,\lambda_{\eta_l}^{(l)}$ are the $\eta_l$ non-zero eigenvalues of ${\bf {F}}^{(l)}$.

With the same approach usually adopted for trellis codes, the
\ac{ST-GTF} in \eqref{T(D)} can now be directly used to evaluate
the error probability as
%
%$$ \tilde{P}_{w}\leq {1\over 2}\sum_{i_1}..\sum_{i_L} w(i_1 ..
%i_L) \EX{{\rm erfc}\sqrt{{E_b \over N_o}{R}
%\sum_{s=1}^{m}\sum_{l=1}^{L}\sum_{i=1}^{\eta_l}\lambda _{i}^{(l)}\left|
%\beta _{i,s}^{(l)}\right| ^{2}}
%%\sum_{l=1}^L i_l H_l^2
% }
%$$
\begin{equation}
\tilde{P}_{w} \leq {1\over 2} \, \EX{ {\cal T}[T_M({\bf
D})]\left|{d_{l,i}={\rm exp}\left(-{E_b \over
N_0}{R}\sum_{s=1}^{m}| \beta _{i,s}^{(l)}| ^{2} \right)} \right.}\,.
%={1\over 2}\sum_{i_1}..\sum_{i_L}
%w(i_1 .. i_L) {\rm exp}\left(-{E_b \over N_o}{R} \sum_{s=1}^{m}\sum_{l=1}^{L}\sum_{i=1}^{\eta_l}\lambda _{i}^{(l)}\left|
%\beta _{i,s}^{(l)}\right| ^{2}\right)\,,
\end{equation}
% where $\beta _{i,s}^{(l)}$ are i.i.d. complex Gaussian r.v.'s.
%
%
This result is due to the well known bound $\erfc(x)\leq e^{-x^2}$
for $x>0$. Tighter bounds and approximations can be obtained by
using the results in \cite{ChiDarSim:J03}, for example with the
exponential bound $\erfc(x)\leq \frac{1}{2} e^{-2 x^2}
+\frac{1}{2} e^{-x^2} < e^{-x^2}$, or with the approximation
$\erfc(x)\simeq \frac{1}{6} e^{-x^2} +\frac{1}{2} e^{-\frac{4 }{3}
x^2}$.
For large \ac{SNR} the asymptotic union bound becomes, as in \eqref{CEP3}:
\begin{equation}
\tilde{P}_{w\infty} \approx  %\frac{1}{4}
K(m \tilde{\eta} _{\min})\ \tilde{F}_{\min}(m)\ \left(\frac{E_s}{4
N_0}\right)^{-\tilde{\eta} _{\min}\cdot m} \,,
\end{equation}
where
\begin{equation}
\tilde{F}_{\min}(m)=
\sum_{({\bf {F}}^{(1)},\ldots,{\bf {F}}^{(L)})\in
I}w\left(
{\bf {F}}^{(1)},\ldots,{\bf {F}}^{(L)}\right) \cdot
\left[ {\prod_{l=1}^{L}%
\prod_{i=1}^{{\eta _{l}}}\lambda _{i}^{(l)}}\right] ^{-m}\,,
\end{equation}
$\lambda _{i}^{(l)}$ are the eigenvalues of ${\bf {F}}^{(l)}$, ${\eta _{l}}$ is the rank of ${\bf {F}}^{(l)}$,
$\tilde{\eta} _{\min }=\min{\sum_l \eta_l}$ and
$$I=\left\{ ({\bf {F}}^{(1)},\ldots,{\bf {F}}^{(L)}):\sum_{l=1}^{L}\eta _{l} =\tilde{\eta}
_{\min }\right\}$$
is the set of error matrices giving $\tilde{\eta} _{\min }$.
%\end{itemize}

We conclude the section with few remarks:

1) The \ac{ST-GTF} depends on both encoder and interleaver
structures, which have been suitably considered to build the error
state diagram specialized to \ac{BFC}.

2) If we are interested in the evaluation of $P_w(\underline{c})$
conditioned to a selected reference codeword $\underline{c}_0$
(usually the one obtained with the all-zero input sequence) we
need to define a \ac{ST-GTF} referred to that sequence
$\underline{c}_0$, which can be easily obtained by using scalar (not
matrix) labels in the error trellis diagram and the modified error
trellis diagram. In this case, the generic transition
$s_{e}^{(t)} \longrightarrow s_{e}^{(t+1)}$ of the error
trellis has to be labeled with
%with a new matrix label $ {\bf E\Delta}(\sigma_e^{(t)},\sigma_e^{(t+1)})$ whose generic element $i,j$ is given by
%
%\begin{equation}
$\Delta_{t}^{{\bf A}^{(t)}}$
%\end{equation}
%
where ${\bf A}^{(t)}$ is the $(n\times n)$ matrix given by ${\bf
A}^{(t)}= \left( \widetilde{{\bf C}}_{s_b^{(t)}\rightarrow
s_b^{(t+1)}}-\widetilde{{\bf C}}_{s_e^{(t)}\oplus
s_b^{(t)}\rightarrow s_e^{(t+1)}\oplus
s_b^{(t+1)}}\right) \left( \widetilde{{\bf
C}}_{s_b^{(t)}\rightarrow s_b^{(t+1)}}-\widetilde{{\bf
C}}_{s_e^{(t)}\oplus s_b^{(t)}\rightarrow
s_e^{(t+1)}\oplus s_b^{(t+1)}} \right)^{H}$ and
$s_b^{(0)},\ldots,s_b^{(N)}$ is the sequence of encoder
states for sequence $\underline{c}_0$ (usually the all-zero state
sequence). Moreover, at each state $s_e^{(t)}=s$ the weighting polynomial
 is a scalar (not a vector), ${ Q}_{s}({\bf D})$, and the \ac{ST-GTF}
is simply obtained as $T_M({\bf D})=  { Q}_{O_N}({\bf D})-1$.

3) In a similar way, if the goal is to find the error probability
according to \eqref{pe_fe} or to a tighter bound obtained by
limiting the set of decoded sequences $\underline{g}$ to those
corresponding to paths in the trellis diagram of code diverging
only once from the path of codeword $\underline{c}$, we can define
a modified error trellis diagram by splitting the all-zero state
at each time $t$, denoted by $O_t$, into two states: $\hat{O}_t$,
having only transitions departing to all the other states
$s_e^{(t+1)} \neq O_{t+1}$, and $\dot{O}_t$, having only the
transition departing  to
$O_{t+1}$ and all the transitions arriving from $s_e^{(t-1)}$. By defining the {\it time-t} \ac{ST-GTF} of this
diagram as $T_{Mt}({\bf D}) \triangleq {\bf Q}_{\dot{O}_N}({\bf D})^T {\bf U}_0$, when
the initial settings are ${\bf Q}_{s_e^{(0)}
\neq \hat{O}_0}({\bf D})=(0,\ldots,0)^T$,
${\bf Q}_{\hat{O}_t}({\bf D})=(1,\ldots,1)^T$ and ${\bf Q}_{\hat{O}_{t'}}({\bf D})=(0,\ldots,0)^T$ for $t'\neq t$, we can obtain: \\
- the $time-0$ \ac{ST-GTF} $T_{M0}({\bf D})$ whose use to evaluate
\eqref{pe_fe}  is
straightforward,\\
- the transfer function $T'_{M}({\bf D})=\sum_{t=0}^{N-1} T_{Mt}({\bf D})$ which can be used in place of $T_{M}({\bf D})$ to refine the
error probability bound.

An example of evaluation of the \ac{ST-GTF} for \ac{P-STC} over
\ac{BFC} is given in Appendix.

%%%%%%%%%%%%%% nuovo commento
\subsection{Discussion on the geometrical uniformity for \ac{STC} and \ac{P-STC}}

Note that the error probability given in \eqref{CEP1}
is in general a function of the reference codeword
$\underline{c}$. The conditions under which there is no dependence
on the transmitted codeword are related to the concept of
geometrical uniformity, that has been introduced in \cite{For:91} with respect
to Euclidean distance.

Geometrically uniform codes are codes with the same distance profile for all pairs of codewords. %where the distance profiles
%from any codeword to all other codewords are equal.
In \ac{AWGN}
channels, the geometrical uniformity
%where the metric is the Euclidean distance (as usually assumed \cite{For:91}),
guarantees that the performance is independent on the particular
transmitted codeword. Thus, the frame error probability can be
evaluated by assuming the transmission of a particular codeword,
that can be the 'all-zero' codeword generated when all the
information (input) bits are $0$. Clearly, this condition greatly
simplifies the code design.

However, the application of the concept of geometrical uniformity
to \ac{STC} requires a careful investigation, as highlighted in
\cite{YanIon:04}. Indeed, it can be noticed that the \ac{PEP}
depends on the Euclidean distance of the coded signals after the
multiple-input multiple-output channel an hence on the eigenvalues
of matrices like those defined in \eqref{eq:Fcg}, that can change
with the reference codeword. For this reason, in general the
design of \ac{STC} should consider all possible transmitted
codewords.

For the \ac{P-STC} introduced in section~\ref{sec:pstc} we can
easily see that:
\begin{itemize}
\item %the codes are geometrical uniform with respect to the
%Euclidean distance. In fact, for the \ac{P-STC} with Gray mapping the
%Euclidean distance between the symbols $c_i^{(t)}, g_i^{(t)}$ of
%two generic codewords, $\underline{c},\underline{g}$, is
%$d_E(c_i^{(t)}, g_i^{(t)})=\sqrt{|c_i^{(t)}-g_i^{(t)}|^2}=2
%\sqrt{d_H(c_i^{(t)},g_i^{(t)})$ for \ac{BPSK}, and
%$d_E(c_i^{(t)},g_i^{(t)})=\sqrt{ |c_i^{(t)}-g_i^{(t)}|^2}=\sqrt{2}
%\sqrt{d_H({c}^I,{g}^I)+d_H({c}^Q,{g}^Q)}}$ for QPSK, where $d_H$
%denotes Hamming distance and $c^I, c^Q, g^I, g^Q $ refer to the
%real and imaginary parts of $c_i^{(t)}$ and $g_i^{(t)}$,
%respectively. Since we are using convolutional codes, the Hamming
%distance spectrum is independent on the reference codeword, and
%therefore the same is true for the Euclidean distance spectrum.
The \ac{P-STC} before the channel (the set of codewords
$\underline{c}$) are geometrically uniform with respect to the
Euclidean distance. In fact, for the \ac{P-STC} with Gray mapping
the Euclidean distance between the symbols of two generic
codewords, $\underline{c},\underline{g}$, is
$d_E(\underline{c},\underline{g})=\sqrt{\sum_t \sum_i
|c_i^{(t)}-g_i^{(t)}|^2}=2
\sqrt{d_H(\underline{c},\underline{g})}$ for \ac{BPSK}, and
$d_E(\underline{c},\underline{g})=\sqrt{\sum_t \sum_i
|c_i^{(t)}-g_i^{(t)}|^2}=\sqrt{2}
\sqrt{d_H(\underline{c}^I,\underline{g}^I)+d_H(\underline{c}^Q,\underline{g}^Q)}$
for \ac{QPSK}, where $d_H$ denotes Hamming distance and the
superscripts $I$, $Q$ refer to the real and imaginary parts,
respectively. Since we are using convolutional codes, the Hamming
distance spectrum is independent of the reference codeword, and
therefore the same is true for the Euclidean distance spectrum.
\item In a system with ideal symbol interleaving
(\ac{BFC} with $L=N$), the \ac{PEP} in \eqref{PairEerfc} depends
only on the Hamming distance between the two codewords, but not on
the specific reference codeword chosen. In fact, the \ac{PEP}
depends on the statistical distribution of the distance after the
channel, defined in \eqref{eq:d2cgZ}. In Rayleigh fading channels,
each $h_{i,s}^{(t)}$ is a complex zero-mean Gaussian distributed
r.v. with variance $1/2$ per dimension; then the generic term
$h_{i,s}^{(t)}\left(c_i^{(t)}-g_i^{(t)}\right)$ is still zero-mean
complex Gaussian with variance $0.5
\left|c_i^{(t)}-g_i^{(t)}\right|^2$. Note that the variance is
thus proportional to the Hamming distance previously discussed
between $c_i^{(t)}$ and $g_i^{(t)}$. The resulting overall
variable $\sum_i h_{i,s}^{(t)}\left(c_i^{(t)}-g_i^{(t)}\right)$ is
still zero-mean complex Gaussian, with a variance that depends
only on the Hamming distance between the codewords $\underline{c}$
and $\underline{g}$. Since for ideal interleaving the \acp{r.v.}
$h_{i,s}^{(t)}$ are \ac{i.i.d.} also in $t$, we can conclude that
the distribution of the r.v. defined in \eqref{eq:d2cgZ} depends
only on the Hamming distance between the codewords.\footnote{It
can be also shown that in \eqref{d2F} the matrix ${{\bf
F}}^{(l)}(\underline{c},\underline{g})={{\bf
A}}^{(l)}(\underline{c},\underline{g})$ has only one non-zero
eigenvalue given by $\lambda_1^{(l)}=\sum_i
|c_i^{(l)}-g_i^{(l)}|^2$ directly related to the Hamming distance
of supersymbols ${\bf C}^{(l)}$ and  ${\bf G}^{(l)}$. } Thus,
since for \ac{P-STC} the Hamming distance spectrum is invariant
with the reference codeword, the same applies to the error
probability bound \eqref{pcod}.
\item In the other cases and especially for quasi-static fading channels
($L=1$), although the code is geometrically uniform before the
channel, we could expect that the \ac{PEP} depends in general on
the reference codeword.
In fact, it depends, through
 \eqref{eq:d2eigen}, on the eigenvalues of the matrix ${{\bf
F}}^{(1)}(\underline{c},\underline{g})$ defined in \eqref{eq:Fcg}.
However, in many cases we have numerically verified that the
conditional error probability does not change significantly
 with the selected reference codeword.
This happens in particular when:\\
- The number of fading blocks $L$ in the \ac{BFC} is large enough
with respect to the length of the error sequences;
in this case the behavior of the ideally interleaved code is approached.\\
- The memory of the code is small, and consequently the error
sequences are short. In this case for many codes
%Since the codes we are investigating must reach the maximum
%diversity, the $\lambda_i^{(1)}$ are all greater than zero;
the distance in \eqref{eq:d2eigen} has a distribution over the set
of all possible codewords $\underline{c}$ which is mainly driven
by the sum of the eigenvalues, i.e.,  the trace of ${{\bf
F}}^{(l)}(\underline{c},\underline{g})$. This is again related to
terms $\left|c_i^{(t)}-g_i^{(t)}\right|^2$ and therefore to the
Hamming distance between the codewords. Thus, the performance is
mainly determined by the Hamming distance spectrum that, in
\ac{P-STC}, is invariant with the reference codeword. This will be
verified
numerically in section~\ref{sec:numres}.\\
Moreover, it is also worth noting that for \ac{P-STC} with $n=2$
antennas and \ac{BPSK} modulation the error probability evaluated
with the all-zero sequence as a reference codeword is always the
worst-case error probability.\footnote{This can be proved (not included here for conciseness) by
looking at the structure of matrix ${{\bf
F}}^{(l)}(\underline{c},\underline{g})$ and its eigenvalues.}

%
% This matrix is the sum of several matrices ${{\bf
%A}}^{(t)}, \,\, t=1,\ldots, N$: for a sufficiently large $N$ we
%can expect to have all possible combinations of super-symbols,
%giving a matrix ${{\bf F}}^{(l)}$ with random elements in a each
% to, ... {\bf TBC}

\end{itemize}

In general we will not rely on the geometrically uniformity
assumption (that holds before the channel but not after the
channel), and so we analyze and design the \ac{P-STC} by averaging
over all possible transmitted sequences.
%
%Furthermore, since for the reasons above we expect that in many
%situations the codes could behave similarly with different
%transmitted codewords, we will also provide some results on the
%design of \ac{P-STC} by assuming the zero reference codeword (so
%simplifying the search).
We will
also show, however, that fixing a particular reference codeword
gives often similar results.

%====================================================

%%%%%%%%%%%%%%%%%%%%%%%%%%%%%%%%%%%%%%%%%%%%%%%%%%%%%%%%%%%%%%%%%%%%%%%%%%%%%%%%
\section{Search for the optimum \ac{P-STC} on \ac{BFC}}\label{sec:search}
%%%%%%%%%%%%%%%%%%%%%%%%%%%%%%%%%%%%%%%%%%%%%%%%%%%%%%%%%%%%%%%%%%%%%%%%%%%%%%%%
%\note{VELIO}

In this section we address the issue of doing an efficient search
of the optimum (in the sense defined later) generators for
\ac{P-STC} in \ac{BFC}.
Our search criterion is based on the asymptotic error probability
in \eqref{CEP3}, so that the optimum code with fixed parameters
$(n,k,h,\mu)$, among the set of non-catastrophic codes, is the
code that

- maximizes the achieved diversity, $\tilde{\eta} _{\min }$;

- minimizes the performance factor $\tilde{F}_{\min}(m)$;

\noindent where the values of $\tilde{\eta} _{\min }$ and
$\tilde{F}_{\min}(m)$ can be extracted from the \ac{ST-GTF} of the
code. Therefore, an exhaustive search algorithm should evaluate the
\ac{ST-GTF} for each code of the set.

Another search criterion for \ac{STC} has been addressed in
\cite{BarBauHan:00,YanBlu:02} where a method based on the
evaluation of the worst \ac{PEP} was proposed. Although the worst
 \ac{PEP} carries information about the achievable
diversity, $\tilde{\eta} _{\min }$, it is incomplete with respect
to coding gain, thus producing a lower bound for the error probability. Even though our
 method based on the union bound is still approximate with
respect to coding gain (giving an upper bound) it includes more
information than the other method, leading often to the choice of codes with better performance.

When applying our search criterion we must consider that, as shown
in \cite{MalLei:99B}, the union bound for the average error
probability is loose and in some cases (long codes and small
diversity) is very far from the actual value. This problem can be
partially overcome by truncating the sum to the most significant
terms, but this technique leads to an approximation. However, this
approach gives good results in reproducing the correct performance
ranking of the codes among those achieving the same diversity
$\tilde{\eta} _{\min }$,  as will be checked in the numerical
results section.

Of course, the achievable diversity is the most important design
parameter. Since $\tilde{\eta} _{\min }$ can not be larger than
both $\eta (\underline{c},\underline{g})\leq n L $ and  the free
distance $d_f$ of the convolutional code used to build the
\ac{P-STC}, it appears that to capture the maximum diversity per
receiving antenna offered by the channel, $nL$, the free distance
of a good code for a given \ac{BFC} should be at least $nL$ or
larger. On the other hand, there is a fundamental limit on  the
achievable diversity related to the Singleton bound for \ac{BFC}
\cite{MalLei:99}. In fact, if we define the \ac{RBFC} for
the system as the ideal equivalent \ac{BFC} with $nL$ fading
blocks that would describe the space-time fading channel if the
$n$ transmitters determine $n$ independent channels, the
achievable diversity, which can not be larger than the diversity
achievable on the reference \ac{BFC}, is bounded by
\begin{equation}
\tilde{\eta} _{\min }\leq 1+\left\lfloor Ln \left(1-{k \over nh}\right) \right\rfloor \,. \label{eq:SB}
\end{equation}
As an example, to achieve full diversity $n$  with a \ac{P-STC} on
a quasi-static channel ($L=1$) the value $k/h$ can not be larger
than $1$, thus the code rate of the convolutional code can not be
larger than $1/n$, or the value of $h$ can not be smaller than $k$
(see also \cite{TarSesCal:98}).

Different methodologies can be used to compute the \ac{ST-GTF} of
the error trellis diagram: we can easily derive an error state
diagram (by splitting all-zero error state) from the related
trellis and in principle use classical techniques to evaluate
$T_{M0}({\bf D})$, but this approach is limited to long codewords
with periodical interleaving since it could be computationally
difficult to handle large matrices. The most efficient method
to compute the \ac{ST-GTF} is to proceed along the error trellis with an
iterative algorithm which evaluates for each state $s_e^{(t)}=s$ the weighting
vector polynomials ${\bf Q}_{s}({\bf D})$  starting from  $t=1$
and ending in $t=N$ with the initial conditions ${\bf Q}_{O_0}({\bf D})=(1,\ldots,1)^T$
and ${\bf Q}_{s_e^{(0)}}({\bf D})=(0,\ldots,0)^T$. This method is also considered in
\cite{ChiConTra:04}. Since $2^{k}N_s$ branches connect the
states of the trellis at each step, there are $2^{k}N_s^3$
products of polynomials (that can be reduced to $2^{2k}N_s^2$
to account for zeros in matrix labels, and further reduced to
$2^{k}N_s$ if labels are scalars).

With the view to utilize the \ac{ST-GTF} to compare different codes in a
systematic search for best codes, the previous algorithm still
maintains a large complexity due to growth of the number of
polynomial terms in the node weights when $t$ increases, and only
conventional simplification rules are available to reduce the
evaluation complexity, which do not allow a significant
improvement in the efficiency of the computation.

This last issue is addressed in \cite{ChiConTra:04} for
convolutional codes over \ac{BFC}, where some simplification rules
are given to largely reduce the computation complexity. Similar rules can be applied to derive the most significant terms of the
\ac{ST-GTF}, namely, those having small diversity order and
product-degree, which allow the evaluation of $\tilde{\eta} _{\min
}$ and $\tilde{F}_{\min}(m)$.

In order to formulate this method let us consider the error state
diagram modified by splitting the all-zero states at each time $t$
and, for each state $s_e^{(t)}=s$, the weighting vector polynomials ${\bf
Q}_{s,r}({\bf D})$ which can be evaluated for the state $s$ by
using the initial settings ${\bf Q}_{s_e^{(0)} \neq \hat{O}_0}({\bf
D})=(0,\ldots,0)^T$, ${\bf Q}_{\hat{O}_{t-r}}({\bf D})=(1,\ldots,1)^T$ and ${\bf
Q}_{\hat{O}_{t'}}({\bf D})=(0,\ldots,0)^T$ for $t'\neq t-r$. Let us also denote
with ${\cal U}_r^{(t)}({\bf D})$ the set of $N_s$ vector
polynomials ${\bf Q}_{S_t,r}({\bf D})$, obtainable for each $s_e^{(t)}=s$ at time $t$. The sets of vector
polynomials ${\cal U}_r^{(t)}({\bf D})$ can be computed along the
error trellis, as for scalar weighting polynomials (see Appendix
of \cite{ChiConTra:04}), with an iterative algorithm which starts
from  $t=1$ and ends at $t=N$  with the required initial setting
for the all-zero state polynomial. We obtain at the end of the
trellis $T'_{M}({\bf D})= \sum_{r=1}^N {\bf Q}_{\dot{O}_N,r}({\bf
D})^T {\bf U}_0$ and, if useful, $T_{M0}({\bf D})=  {\bf Q}_{\dot{O}_N,N}({\bf
D})^T {\bf U}_0$.

The evaluation of the codeword error probability through the
iterative computation for each $r$ of the sequence ${\cal
U}_r^{(1)}({\bf D}),...,{\cal U}_r^{(N)}({\bf D})$ leads to the
possibility of setting up much more efficient computation of a
truncated asymptotic bound. To this aim we use the two following
properties of non-negative definite Hermitian matrices
\cite{HorJoh:B99}:

P1) the rank of the sum of two
non-negative definite Hermitian matrices is greater than, or
equal to, the rank of each matrix;

P2) the product of non-zero eigenvalues of the sum of two
non-negative definite Hermitian matrices is greater than, or equal
to, the product of non-zero eigenvalues of each matrix.
%P) the (ordered) eigenvalues of the sum of two
%non-negative definite Hermitian matrices are greater than, or
%equal to, the eigenvalues of each matrix.

\noindent Then, additional simplification rules are possible in
order to eliminate polynomial terms which do not affect the final
value of
$\tilde{\eta} _{\min }$ and  $\tilde{F}_{\min}(m)$). In fact, by means of P1 and P2, respectively, at each step $t$ it is possible:\\
- to eliminate from each element of ${\bf Q}_{s,r}({\bf D})$ the
polynomial terms with rank of the exponent strictly greater than
the minimum rank of the exponent of the polynomial
terms in ${\bf Q}_{O,r}({\bf
D})$;\\
- to eliminate  from each element of vector ${\bf Q}_{s,r}({\bf
D})$ the polynomial terms with product of non-zero eigenvalues of the exponent
much greater (a threshold should be fixed) than the
minimum product of non-zero eigenvalues  of the polynomial terms with
minimum rank  in ${\bf Q}_{O,r}({\bf
D})$. %(this rule
%may be carefully used in the more computationally hard cases).

%%%%%%%%%%%%%%%%%%%%%%%%%%%%%%%%%%%%%%%%%%%%%%%%%%%%%%%%%%%%%%%%%%%%%%%%%%%%%%%%
\section{Numerical results}
%%%%%%%%%%%%%%%%%%%%%%%%%%%%%%%%%%%%%%%%%%%%%%%%%%%%%%%%%%%%%%%%%%%%%%%%%%%%%%%%
\label{sec:numres}
%\note{ANDREA}
This section presents the results of the previously proposed
search algorithm used to design \ac{P-STC} in \ac{BFC}, and their
performance (by simulation) in terms of \ac{FER} versus the
\ac{SNR} defined as $E_b/N_0$ per receiving antenna element. In
addition, comparisons with the performance of previously known
\acp{STC} are also given. All simulations are performed with
random generation of information bits, thus without fixing a reference transmitted codeword, and with MIMO($n$,$m$) we refer to a system with $n$ transmit antennas and $m$ receive antennas.

First we investigate how the \ac{P-STC} architecture exploits the
diversity in \ac{BFC}.
To this aim, we evaluate the suitability of the pragmatic approach
considering some \ac{P-STC} obtained using the known optimal
convolutional code designed for the \ac{AWGN} channel. As an
example, for \ac{BPSK} modulation with $n$ transmitting antennas,
a rate $1/n$ binary code is used, with a spectral efficiency of $1
bps/Hz$.

\begin{figure}
\centerline{\includegraphics[width=.95\linewidth,draft=false]{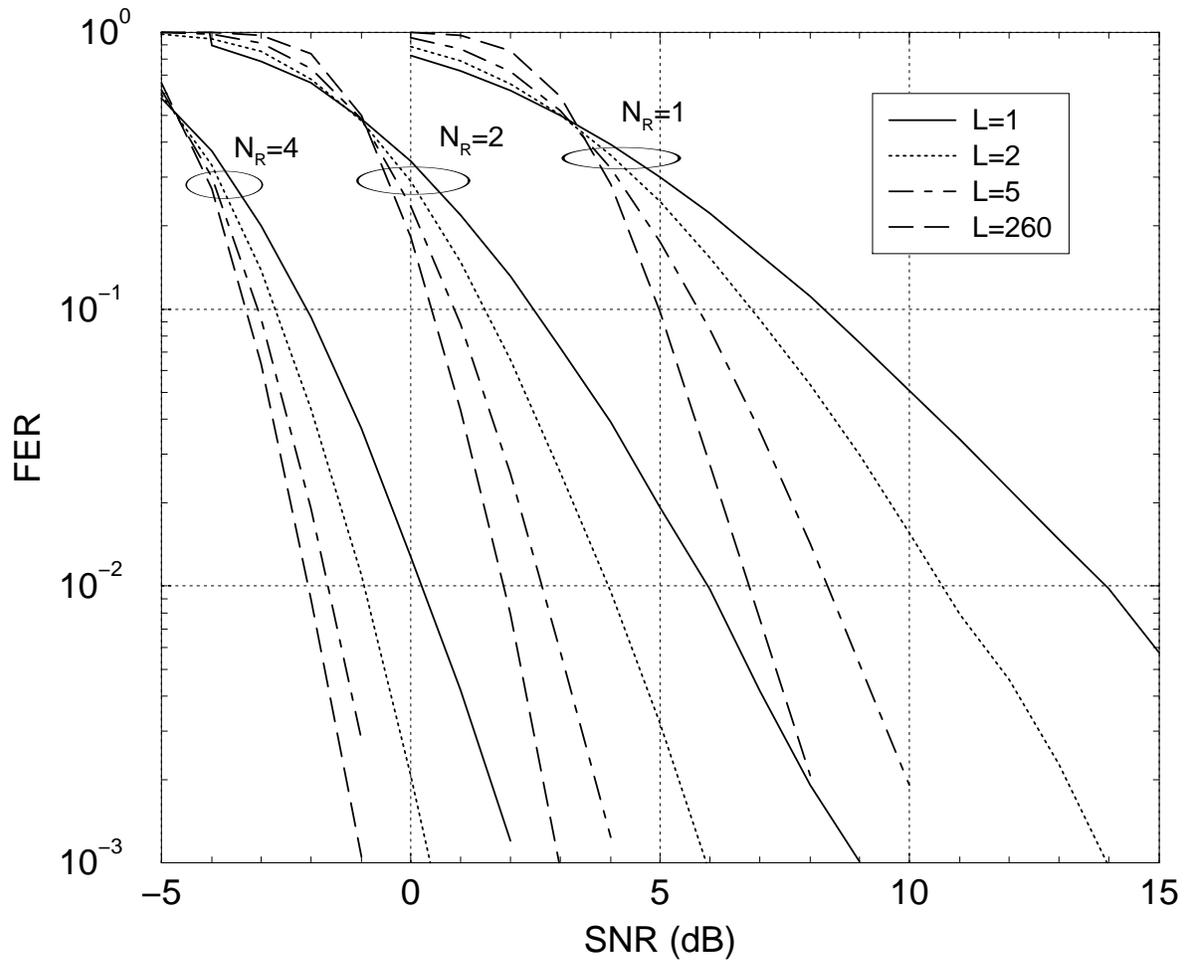}}
\caption{FER vs. SNR for \ac{P-STC} obtained with the rate $1/2$,
$64$ states convolutional encoder with generators $(133,171)_8$,
$1 \, bps/Hz$ \ac{BPSK}, $n=2$ transmitting antennas and $m=1,2,4$
receiving antennas in \ac{BFC} for various $L$.}
\label{fig:bpsk_1_2_7_2xNr_130_L}
\end{figure}

In Fig.~\ref{fig:bpsk_1_2_7_2xNr_130_L} the \ac{FER}
 for a \ac{BPSK} modulated \ac{P-STC} obtained
with the de-facto standard $64$ states convolutional encoder with
octal generators $(133,171)_8$ is shown, assuming a \ac{BFC}. In
particular, $n=2$ transmitting antennas and $m=1,2,4$ receiving
antennas are considered for various fading rates given by
$L=1,2,5,260$ fading levels per codeword with $N=130$. Note that
even with this not-optimized choice of the generators, \ac{P-STC}
are able to reach the maximum achievable spatial diversity; in particular
for $L=1$ (i.e., quasi-static fading channel, meaning absence of
time diversity), the diversity order is given by the product $n\cdot
m$. For  $L$ greater than $1$, thus in the presence of available
time-diversity, the achieved diversity order increases, depending
also on the number of states. For MIMO(2,2) \ac{P-STC}, typical
values of interest for the \ac{FER} (i.e., in the order of
$10^{-2}$) can be reached with $E_b/N_0$ per receiving antenna
element of about $6$ dB (quasi-static channel), $3.8$ dB ($L=2$),
$2.5$ dB ($L=5$) and $2$ dB (fully-interleaved case), whereas with
$4$ receiving antennas the required \ac{SNR} decreases to $0.2$
dB, $-1.1$ dB, $-1.6$ dB and $-2.1$ dB, respectively.
%\footnote{We would like also compare these results with the literature, but at the authors knowledge no results of other STCs with \ac{BPSK} are available for comparison.}

\begin{figure}
\centerline{\includegraphics[width=.95\linewidth,draft=false]{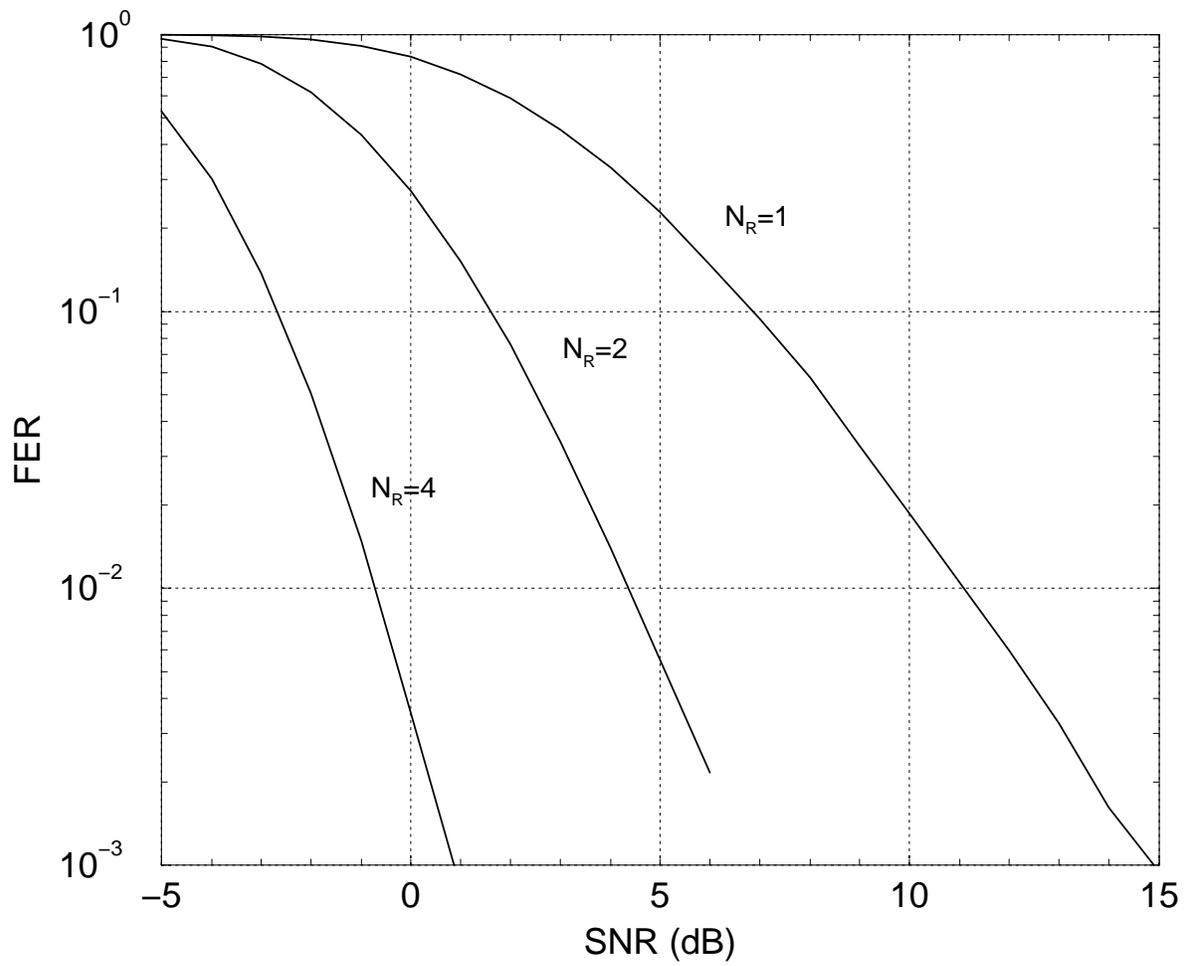}}
\caption{FER vs. SNR for \ac{P-STC} obtained with $1/4$
convolutional encoder, $64$ states, \ac{BPSK}, $1\ bps/Hz$, $4$
transmitting antennas and $1,2,4$ receiving antennas in
quasi-static Rayleigh fading channel.}
\label{fig:bpsk_1_4_7_4xNR_130_L1}
\end{figure}

The low complexity of the \ac{P-STC} architecture makes also
feasible the use of a larger number of transmitting antennas. As
an example, the case of $n=4$ is shown in
Fig.~\ref{fig:bpsk_1_4_7_4xNR_130_L1} for quasi-static Rayleigh
fading, $N=130$ and $m=1,2,4$. Here, the convolutional encoder
with optimal \ac{AWGN} generators $(135,135,147,163)_8$ is adopted
\cite{Pro:B01}. Note that the case MIMO(4,2) achieves \ac{FER}
equal to $10^{-2}$ at $4.2$ dB, that is greater than the $0.2$ dB
of the case MIMO(2,4) seen before; this is
 due to both  the different power repartition on transmitting antennas and power combining at receiving antennas as well as the different
code-rate. % (note that this would apply even considering the total SNR).

Similarly, by using the generators for \ac{CC} over \ac{AWGN} it
is possible to design $n$-\ac{P-STC} for \ac{QPSK} ($2$ bps/Hz) by
using the rate $2/2n$ convolutional codes.
%The simpler way to design it is
%to use two rate $1/n$ encoders, in parallel. Then, in order to
%exploit diversity, the outputs of each encoder are split to
%different antenna. The number of states of the resulting encoder
%is the product of that of the constituent encoders.
%For example, in Fig. \ref{pstcL1st64bpsk} we show the performance of a 64 states \ac{P-STC}, for $n=2$, QPSK, obtained by two, rate $1/2$ convolutional encoders with 8 states and polynomials generators $(15,17)_8$ optimum for \ac{AWGN}. The resulting $2/4$, 64 states encoder has generators $(242,121,252,125)_8$.

%\vskip 2cm \note{Velio hai tabelle da inserire su ricerca codici? Se si le metterei qui.}\vskip 2cm

Let us now consider the search for optimum generators (in the sense defined
in section~\ref{sec:search}). %, for the quasi-static
%fading channel. for which we can perform a code search following
%the procedure outlined in the previous section.
In Tab.~\ref{tab:2/4,mu2,rx2} we report, for the quasi-static
fading channel and \ac{QPSK}, the characteristic parameters and
performance of the best generators for the $(4,2,2) 2$-\ac{P-STC},
compared with the code proposed in \cite{TarSesCal:98}
\footnote{It is possible to show that for these parameters the
code given in \cite{TarSesCal:98} is amenable of a \ac{P-STC}
representation with generators $(01,02,04,10)_8$.}. Note that all
codes achieve the available diversity $n\cdot m$, but with different
performance factors $\tilde{F}_{\min}(m)/N$. It is remarkable that
the ratio between performance factors is almost the same as the
ratio of the simulated \acp{FER}. Moreover, even at SNR of $15$ dB, the
asymptotic bound is sufficiently close to \ac{FER}. Note also that
the performance factor evaluated by fixing a reference codeword
(${F}_{\min}(\underline{c}_0,m)/N$) provides a slightly different
ranking of generators, giving as best code the generator
$(06,13,11,16)_8$. This code is not the best according to
$\tilde{F}_\text{min}(m)$ (that would give $(05,11,06,16)_8$), but
is very close to it terms of performance factor and the best in
terms of \ac{FER}. As will be clarified in the following, we
checked that there are $11$ codes out of $2^{16}$ behaving as the first and $47$
behaving as the second, meaning that there is not a single best
code but several codes that perform similarly.

This fact suggests us to carefully investigate the performance
differences among generators through exhaustive simulations. Thus,
we performed an exhaustive simulation for all possible 4 states
$n=2, m=2$ \ac{P-STC} in terms of \ac{FER} for \ac{QPSK} in
quasi-static fading channel, with $E_b/N_0=9$ dB. In
Fig.~\ref{fig:allcodes9dB} we report the \ac{FER} for all
$4$-states \ac{P-STC} obtained through $2/4$ convolutional
encoders (i.e., $2^{16}$ generators that are ordered in abscissa).
A remarkable outcome is that also for \ac{P-STC} it is verified a
phenomenon similar to what already discussed in
\cite{ChiConTra:04} for convolutional codes in \ac{BFC}:
non-catastrophic codes can be divided in few classes, with almost
the same performance for codes in the same class. Note that within
the class of codes providing the best performance, there is the
one obtained through our searching methodology, that gives a
\ac{FER} of about $0.01$. Even for this simple case of $4$ states
generators, the exhaustive search by simulation required one
entire week on a Pentium 4 personal computer, whereas with our
code searching algorithm we saved about two order of magnitude in
time. An exhaustive search for a larger number of states is
impractical, while our search algorithm works still well,
emphasizing the importance of algorithmic methods.

Hence, it is important to note that the pragmatic structure is not
only interesting from the implementation point of view, but it
also provides interesting performance, that, in all cases we
investigated, outperformed the previously known \acp{STC}.
In order to make the comparison between \ac{P-STC}  and \ac{STC}
possible, in the following numerical results we assume $N=130$
 \cite{TarSesCal:98,BarBauHan:00,YanBlu:02}. As an example, the
performance of our best MIMO(2,2) \ac{QPSK} (4,2,2) 2-\ac{P-STC} (with generators $(06,13,11,16)_8$) is compared in
Fig.~\ref{fig:QPSK_2x2_4stati_2bpsHz_L1_confronto} with other
\ac{STC} known in the literature for quasi-static channel
\cite{TarSesCal:98,BarBauHan:00,YanBlu:02}. These results show
that out \ac{P-STC} outperform previously known \ac{STC}.

Moreover, the proposed code search methodology also enables to
find \ac{P-STC} for various fading rates, that is, for various
values of the parameter $L$. As an example, in the case of MIMO(2,2) \ac{QPSK} (4,2,2) 2-\ac{P-STC}, we obtain that the best
generator is $(05,06,13,17)_8$ when $L \ge 2$, as shown in
Fig.~\ref{fig:QPSK_L_2x2_4stati_2bpsHz} for $L=1,2,5$, and $260$.
At the author knowledge no other results for the \ac{BFC} are
present in the literature, so we compare our codes with the
original \ac{STC} in \cite{TarSesCal:98} even if the latter was
designed for the quasi-static case. Note that with only $4$ states
codes are not able to exploit all available time diversity, but
 the proposed codes already achieve the available spatial
diversity.

Then, we investigate the impact of the number of states on the
performance of MIMO(2,1) \ac{P-STC} with \ac{BPSK} in
quasi-static fading channel. In Tab.~\ref{tab:1/2,L=1,rx1} we
report best codes obtained through the search algorithm for
$2,4,8,16$ states, for which we indicate the achieved diversity,
the performance factor and the \ac{FER}. We also report the
performance factor for \ac{AWGN} optimal generators
with the same number of states. Note that all codes achieve the
maximum diversity, and that increasing the number of states does
not produce relevant performance improvements. Moreover, on the
quasi-static channel the error probability bound tends to become
looser, especially when the free distance of the convolutional
code increases with respect to the achieved diversity.

The behavior is different in \ac{BFC} with time diversity
available. This case is illustrated in Tab.~\ref{tab:1/2,L=8,rx1}
for $L=8$, where it is shown that increasing the number of states
results in a larger diversity. Note also that the optimum \ac{P-STC} are
able to achieve a diversity equal to that achieved
by using the optimal generators for the \ac{AWGN} channel, and
that are not able to reach the diversity achievable on the
\ac{RBFC} with $nL$ fading levels per codeword. This
means that convolutional codes are more capable to collect time
diversity than spatial diversity.

Finally, we report in Tabs.~\ref{tx2,L=1,rx1}, \ref{tx3,L=1,rx1},
and \ref{tx4,L=1,rx1} the optimum generators obtained through the
search algorithm for $n=2,3,4,$ respectively, with \ac{BPSK} and
\ac{QPSK} modulations, and for different number of states. The
corresponding performance factors are also reported. It is worth
noting that, although the codes are not geometrically uniform, in
most cases the code search based on the \ac{ST-GTF} with a fixed
reference codeword leads to the same code as the search over all
possible transmitted codewords, or to a code with similar
performance.
%

%%%%%%%%%%%%%%%%%%%%%%%%%%%%%%%%%%%%%%%%%%%%%%%%%%%%%%%%%%%%%%%%%%%%%%%%%%%%%%%%
\section{Conclusions}
%%%%%%%%%%%%%%%%%%%%%%%%%%%%%%%%%%%%%%%%%%%%%%%%%%%%%%%%%%%%%%%%%%%%%%%%%%%%%%%%
%\note{ANDREA (+BIBLIO+FUGURE)}
In this paper we investigated the feasibility of a pragmatic
approach to \acl{STC}, where common convolutional encoders and
decoders are used, with suitably defined branch metrics.

We extended to \acl{P-STC} the concept of \acl{GTF} for \acl{CC}
in \acl{BFC}, that results in the possibility to rank different
codes with an efficient algorithm, based on the asymptotic error
probability union bound. A search methodology to obtain optimum
generators for different fading rates has then been proposed.

It has been shown that \ac{P-STC} achieve better performance
compared to previously known \ac{STC} and that they are suitable
for systems with different spectral efficiencies, number of
antennas and fading rates, %, taken into account by the \ac{BFC} model.
and are therefore a valuable choice both in terms of
implementation complexity and performance.

%%%%%%%%%%%%%%%%%%%%%%%%%%%%%%%%%%%%%%%%%%%%%%%%%%%%%%%%%%%%%%%%%%%%%%%%%%%%%%%%
\bibliographystyle{IEEEtran}
\bibliography{IEEEabrv,BibMIMO,BiblioMCCV,BibBooks}
%%%%%%%%%%%%%%%%%%%%%%%%%%%%%%%%%%%%%%%%%%%%%%%%%%%%%%%%%%%%%

%\newpage

\section*{Appendix}
\label{sec:app} 
\begin{figure}
%{\small
%\vskip -4cm
%\centerline{\input{appendixPSTC.tex}}}
\centerline{\includegraphics[width=.5\linewidth,draft=false]{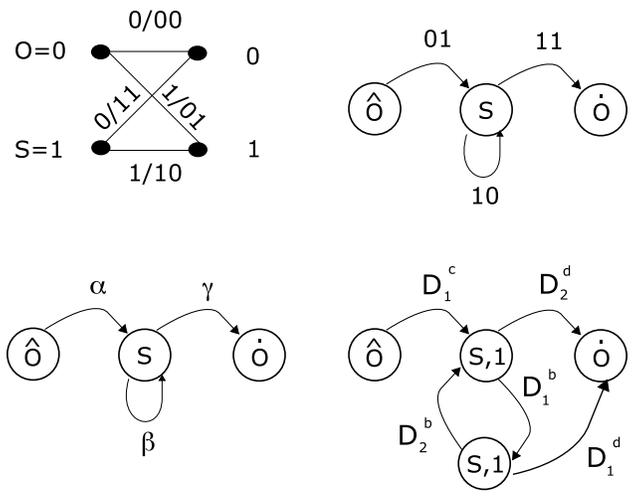}}
%\vskip -8cm
\caption{Trellis and state diagrams for the \ac{P-STC} investigated
in appendix.} \label{fig:appendixPSTC}
\end{figure}

An example of computation of the \ac{ST-GTF} for
\ac{P-STC} is reported here. We consider a (2,1,2) 2-PSTC used
with $m=1$ receiving antenna over a quasi-static \ac{BFC} and
obtained from a $1/2$ convolutional code with generators $(1,3)_8$
and \ac{BPSK} modulation format. These generators are the best for
two-states \ac{CC} in \ac{AWGN}, with free distance 3; when used
to build a \ac{P-STC} we checked that this choice of generators
produces the second best code in quasi-static fading channels,
achieving diversity 2, i.e. the maximum available diversity.
This code %is not able to achieve $\tilde{\eta}_{min}>2$ for
%\ac{BFC} with $L>1$ with any interleaving scheme and it
has been chosen since simple enough to allow the evaluation of
$time-0$ \ac{ST-GTF}, $T_{M0}({\bf D})$, using standard algorithm
on the modified error state diagram. % by splitting the all zero-state (note that the resulting time-$0$ GTF refers to the case with $N$ approaching infinity).
The two-states trellis is depicted in Fig.~\ref{fig:appendixPSTC} (up-left). %, the trellis has two states $O$ and $S$ with transitions: $O\rightarrow O$ for input $0$ with output $00$, $O\rightarrow S$ for input $1$ with output $01$, $S\rightarrow O$ for input $0$ with output $11$, $S\rightarrow S$ for input $1$ with output $10$.

The associated possible output symbols $[c_1; c_2]$ and $[g_1;
g_2]$ are in the set $\{X_0,X_1,X_2,X_3\}$ with $X_0=[-1;-1]$,
$X_1=[-1;1]$, $X_2=[1;-1]$ and $X_3=[1;1]$.  Thus, the matrix
${\bf A}(\underline{c},\underline{g})$ is in the set $\{{\bf
a,b,c,d,e}\}$ with %${\bf a}=[0,0;0,0]$, ${\bf b}=[4,0;0,0]$, ${\bf
%c}=[0,0;0,4]$, ${\bf d}=[4,4;4,4]$, ${\bf e}=[4,-4;-4,4]$
%
$${\bf a}=
  \begin{pmatrix}
    0 & 0 \\
    0 & 0
  \end{pmatrix} \qquad
{\bf b}=%[4,0;0,0]
  \begin{pmatrix}
    4 & 0 \\
    0 & 0
  \end{pmatrix} \qquad
{\bf c}=%[0,0;0,4]
  \begin{pmatrix}
    0 & 0 \\
    0 & 4
  \end{pmatrix} \qquad
{\bf d}=%[4,4;4,4]
  \begin{pmatrix}
    4 & 4 \\
    4 & 4
  \end{pmatrix} \qquad
{\bf e}=%[4,-4;-4,4]
  \begin{pmatrix}
    4 & -4 \\
    -4 & 4
  \end{pmatrix}.
$$%\footnote{For compactness we used notation $[]$ for matrices,
%with commas separating elements in the same row and
% semicolon separating rows.}
The error state diagram modified by
splitting the all-zero state with different labeling $\hat{0}$ and
$\dot{0}$ is given in Fig.~\ref{fig:appendixPSTC} (up-right).
%,by: $\hat{0} \rightarrow S$ with output $01$, $S
%\rightarrow S$ with output $10$, $S \rightarrow \dot{0}$ with
%output $11$.
Thus, by following the steps in Sec.\ref{sec:STGTF} we rewrite the
error state diagram for quasi-static fading channel as in
Fig.~\ref{fig:appendixPSTC} (down-left) %: $\hat{0} \rightarrow S$
%with output $\alpha$, $S \rightarrow S$ with output $\beta$, $S
%\rightarrow \dot{0}$ with output $\gamma$,
where %${\bf \alpha}(D)=1/2\ [D^{\bf c},D^{\bf c};D^{\bf c},D^{\bf
%c}]$, ${\bf \beta}(D)=1/2\ [D^{\bf b},D^{\bf b};D^{\bf b},D^{\bf
%b}]$ and ${\bf \gamma}(D)=1/2\ [D^{\bf d},D^{\bf e};D^{\bf
%d},D^{\bf e}]$.
$${\bf \alpha}(D)=1/2 %\ [D^{\bf c},D^{\bf c};D^{\bf c},D^{\bf c}]
  \begin{pmatrix}
    D^{\bf c} & D^{\bf c} \\
    D^{\bf c} & D^{\bf c}
  \end{pmatrix} \qquad
{\bf \beta}(D)=1/2 %\ [D^{\bf b},D^{\bf b};D^{\bf b},D^{\bf b}]$
  \begin{pmatrix}
    D^{\bf b} & D^{\bf b} \\
    D^{\bf b} & D^{\bf b}
  \end{pmatrix} \qquad
{\bf \gamma}(D)=1/2 %\ [D^{\bf d},D^{\bf e};D^{\bf d},D^{\bf e}]$
  \begin{pmatrix}
    D^{\bf d} & D^{\bf e} \\
    D^{\bf d} & D^{\bf e}
  \end{pmatrix}.
$$

The corresponding \ac{ST-GTF} results in
\begin{equation}
\begin{cases}
   {\bf Q}_S^T(D)={\bf U}_0^T {\bf \alpha}(D) +{\bf Q}_S^T {\bf \beta}(D)  & \\
    {\bf Q}_{\dot{0}}^T(D)= {\bf Q}_S^T(D) {\bf \gamma}(D) & \\
\end{cases}
\Rightarrow
\begin{cases}
   {\bf Q}_S^T(D)={\bf U}_0^T {\bf \alpha}(D) \left({\bf I}- {\bf \beta}(D)\right)^{-1}  & \\
    {\bf Q}_{\dot{0}}^T(D)= {\bf U}_0^T {\bf \alpha}(D) \left({\bf I}- {\bf \beta}(D)\right)^{-1} {\bf \gamma}(D) & \\
\end{cases}
%\nonumber
\end{equation}
giving
\begin{equation}
T_{M0}(D)= {\bf Q}_{\dot{0}}^T(D) {\bf U}_0=\frac{1}{4} [D^{\bf
c}\ D^{\bf c}] \left({\bf I}- {\bf \beta}(D)\right)^{-1} [D^{\bf
d}\ D^{\bf d}]^T=\frac{D^{{\bf c}+{\bf d}}}{1-D^{\bf b}} \,,
\label{eq:appTM0}
\end{equation}
which can also be expanded in
\begin{eqnarray}
T_{M0}(D)&=&D^{{\bf c}+{\bf d}}(1+D^b/(1-D^{\bf b}))= D^{{\bf c}+{\bf d}}+D^{{\bf c}+{\bf d}+{\bf b}}(1+D^{\bf b}/(1-D^{\bf b}))=\ldots \nonumber \\
&=& D^{{\bf c}+{\bf d}}+D^{{\bf c}+{\bf d}+{\bf b}}+D^{{\bf
c}+{\bf d}+2{\bf b}}+\ldots \,.
\end{eqnarray}
Therefore, to obtain the \ac{ST-GTF} of the code we need to
compute the eigenvalues of matrices of the form ${\bf c}+{\bf
d}+i{\bf b}$ with $i\ge0$ integer, and
\begin{equation}
\mathcal{T}\left[T_{M0}(D)\right]=\sum_{i=0}^{+\infty} d_1^{\lambda_{i1}} d_2^{\lambda_{i2}}\,,
\end{equation}
where $\lambda_{i1}$ and $\lambda_{i2}$ are the two eigenvalues of
${\bf c}+{\bf d}+i{\bf b}$ (e.g., $\lambda_{01}=\lambda_{02}=6 \pm
\sqrt{20}$) while their product is simply $\lambda_{i1}
\lambda_{i2}=2i+1$. Then, we obtain $\tilde{\eta}_{\text{min}}=2$
and $\tilde{F}_{\text{min}}(1) \approx (N/16) (1+1/3+1/5+\ldots)$.
It is interesting to note that both ${\bf \alpha}(D)$ and ${\bf
\beta}(D)$ have equal elements, that ${\bf e}$ and ${\bf e}$ can
be interchanged without altering the eigenvalues, and thus the
\ac{ST-GTF} does not depend on the particular reference sequence.
Therefore this code is geometrically uniform even at the output of
the channel. As check we can evaluate $T_0(D)$ for the all-zero
reference codeword $\underline{c}=\underline{c}_0$ by exploiting
the error state diagram with scalar labels; to this aim it is
sufficient to replace $\alpha(D)=D^{\bf c}$, $\beta(D)=D^{\bf b}$
and $\gamma(D)=D^{\bf d}$, so obtaining
\begin{equation}
T_0(D)=\alpha(D) (1-\beta(D))^{-1}\gamma(D)=\frac{D^{{\bf c}+{\bf
d}}}{1-D^{\bf b}}\,,
\end{equation}
which is equal to $T_{M0}(D)$ in \eqref{eq:appTM0}.

To evaluate the \ac{ST-GTF} of the \ac{P-STC} in a \ac{BFC} with
$L=2$ and periodical interleaving, i.e., $\mathbf{
\Delta}=\left(D_1 D_2 D_1 D_2 \ldots\right)$, we can extend the
error state diagram by replacing the state $S$ with two states
$S_1$ and $S_2$, thus obtaining transitions $\hat{0} \rightarrow
S_1$ with output $D_1^{\bf c}$, $S_1 \rightarrow S_2$ with output
$D_2^{\bf b}$, $S_2 \rightarrow S_1$ with output $D_1^{\bf b}$,
$S_1 \rightarrow \dot{0}$ with output $D_2^{\bf d}$ and $S_2
\rightarrow \dot{0}$ with output $D_1^{\bf d}$ (see Fig.
\ref{fig:appendixPSTC} down-right). The \ac{ST-GTF} is then
\begin{equation}
\begin{cases}
    Q_{S_1}(D_1,D_2)=D_1^{\bf c}+Q_{S_2}(D_1,D_2) D_1^{\bf b}  & \\
    Q_{S_2}(D_1,D_2)=Q_{S_1}(D_1,D_2) D_2^{\bf b}  & \\
    Q_{\dot{0}}(D_1,D_2)= Q_{S_1}(D_1,D_2) D_2^{\bf d} + Q_{S_2}(D_1,D_2) D_1^{\bf d} & \\
\end{cases}
\Rightarrow
\begin{cases}
    Q_{S_1}(D_1,D_2)=\frac{D_1^{\bf c}}{1-D_1^{\bf b} D_2^{\bf b}}  & \\
    Q_{S_2}(D_1,D_2)=\frac{D_1^{\bf c} D_2^b}{1-D_1^{\bf b} D_2^{\bf b}}  & \\
    Q_{\dot{0}}(D_1,D_2)= \frac{D_1^{\bf c} D_2^{\bf b} + D_1^{{\bf c}+{\bf d}} D_2^{\bf b}}{1-D_1^{\bf b} D_2^{\bf b}} & \\
\end{cases}
%\nonumber
\end{equation}
and after expansion
\begin{equation}
T_0(D_1,D_2)=T_{M0}(D_1,D_2)=D_1^{\bf c} D_2^d+D_1^{{\bf c}+{\bf
d}} D_2^{\bf b}+D_1^{{\bf c}+{\bf b}} D_2^{{\bf d}+{\bf b}} +
D_1^{{\bf c}+{\bf d}+{\bf b}} D_2^{2{\bf b}}+\ldots \,,
\end{equation}
in which only $D_1^{\bf c} D_2^{\bf d}$ has two terms with rank
$1$. Therefore $\mathcal{T}[D_1^{\bf c} D_2^{\bf d}]=d_{11}^1
d_{21}^1$ implying $\tilde{\eta}_{\text{min}}=2$ and
$\tilde{F}_{\text{min}}(1) \approx N/16$.

\clearpage

\clearpage
\clearpage
\begin{figure}
{\small \centerline{\input{pstc_txbpsk.pic}}} \vskip -16cm
\caption{Example of $1\ bps/Hz$ pragmatic space-time encoder for
$n=2$ transmitting antennas and \ac{BPSK} modulation, obtained
with a rate $1/2$ convolutional encoder with $4$ states and
generator polynomials $(5,7)_8$.} \label{fig:schemePSTCtxbpsk}
\end{figure}
%\clearpage
%%%%%%%%%%%%%%%%%%%%%%%%%%%%%%%%%%%%%%%%%%%%%%%
\begin{figure}
\centerline{\includegraphics[width=.2\linewidth,draft=false]{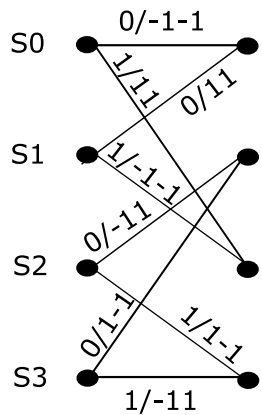}}
\caption{Trellis for the pragmatic space-time code of
Fig.~\ref{fig:schemePSTCtxbpsk}. On each branch the first is the
input bit, followed by the two output antipodal symbols.}
\label{fig:schemePSTCtxbpsktrellis}
\end{figure}
%%%%%%%%%%%%%%%%%%%%%%%%%%%%%%%%%%%%%%%%%
\clearpage
\clearpage

\clearpage

\clearpage

\clearpage

\clearpage
{
\begin{table}
\caption{Comparison of rate $2/4$ \ac{P-STC} with QPSK, $n=m=2$,
$\mu=2$ on \ac{BFC} with $L=1$, $E_b/N_0=15 dB$ and $N=130$. The
performance factor is truncated with $\delta_H=9$. The first two codes are the best produced by
the search (there are 12 first-class codes with almost the same
behavior, and 48 second-class codes). The third code is the code
proposed in \cite{TarSesCal:98}.} \center{
\begin{tabular}{|c|c|c|c|c|c|c|}
  % after \\: \hline or \cline{col1-col2} \cline{col3-col4} ...
  \hline
  {\bf Generators} &  $d_f$ & $\tilde{\eta}_{min}m$ & $\tilde{F}_{\min}(2)/N$ & ${F}_{\min}(\underline{c}_0,2)/N$ & $\tilde{P}_{w\infty}$ & $FER$ \\
  \hline
  $(06,13,11,16)_8$ & 4 &4 & 0.076 & 0.048 & $3.5\ 10^{-4}$ & $9.3\ 10^{-5}$\\
  $(05,11,06,16)_8$ & 4 &4 & 0.073 & 0.092 & $3.4\ 10^{-4}$ & $1.0\ 10^{-4}$\\
  $(01,02,04,10)_8$ & 2 &4 & 0.125 & 0.125 & $5.7\ 10^{-4}$ & $2.4\ 10^{-4}$\\
 \hline
\end{tabular}
}\label{tab:2/4,mu2,rx2}
\end{table}

%\begin{figure}
%\centerline{\includegraphics[width=.8\linewidth,draft=false]{Q4st_L1_2x2_allcodes2.eps}}
%\caption{Exhaustive searching for $2 \times 2$ PSTC in terms of
%FER: QPSK, quasi-static fading channel ($L=1$),  $E_b/N_0=7$dB.}
%\label{fig:allcodes}
%\end{figure}
\begin{figure}
\centerline{\includegraphics[width=.95\linewidth,draft=false]{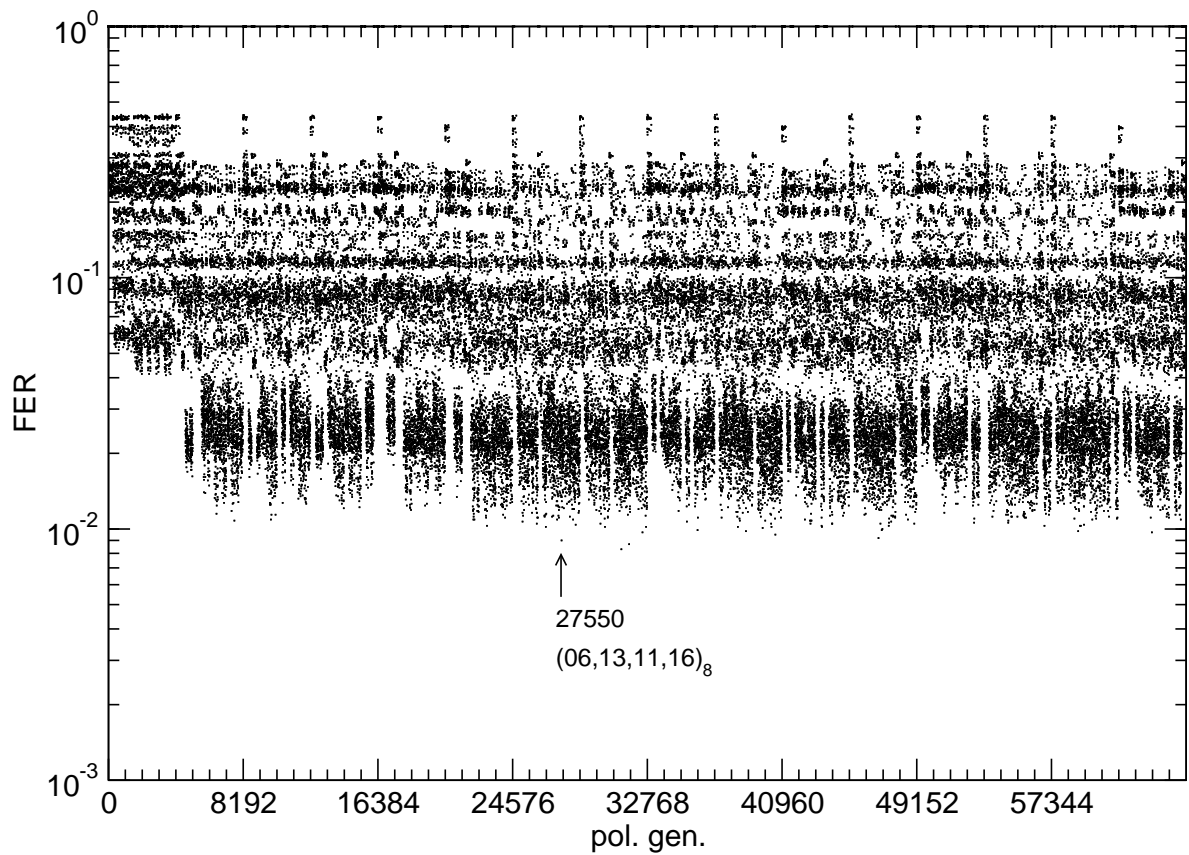}}
\caption{Exhaustive search for MIMO(2,2) \ac{P-STC} in terms of FER:
QPSK, quasi-static fading channel ($L=1$),  $E_b/N_0=9$dB.}
\label{fig:allcodes9dB}
\end{figure}
\begin{figure}
\centerline{\includegraphics[width=.95\linewidth,draft=false]{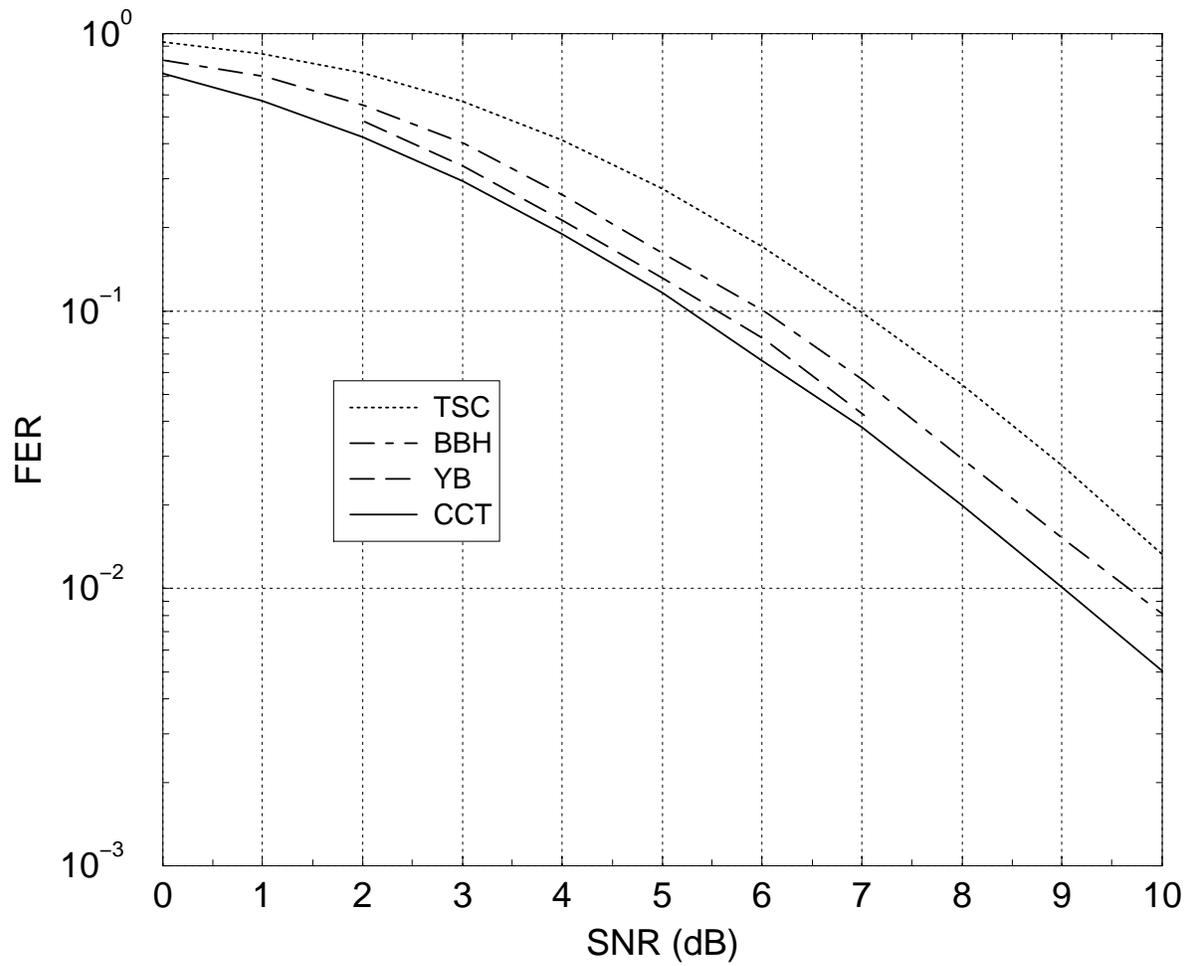}}
\caption{Comparison of our $4$ states QPSK MIMO(2,2), $2bps/Hz$, \ac{P-STC}
(continuous line) and previously known \acl{STC}. TSC:
\cite{TarSesCal:98}, BBH: \cite{BarBauHan:00}, and YB:
\cite{YanBlu:02}, quasi-static fading channel.}
\label{fig:QPSK_2x2_4stati_2bpsHz_L1_confronto}
\end{figure}

\clearpage

\begin{figure}
\centerline{\includegraphics[width=.95\linewidth,draft=false]{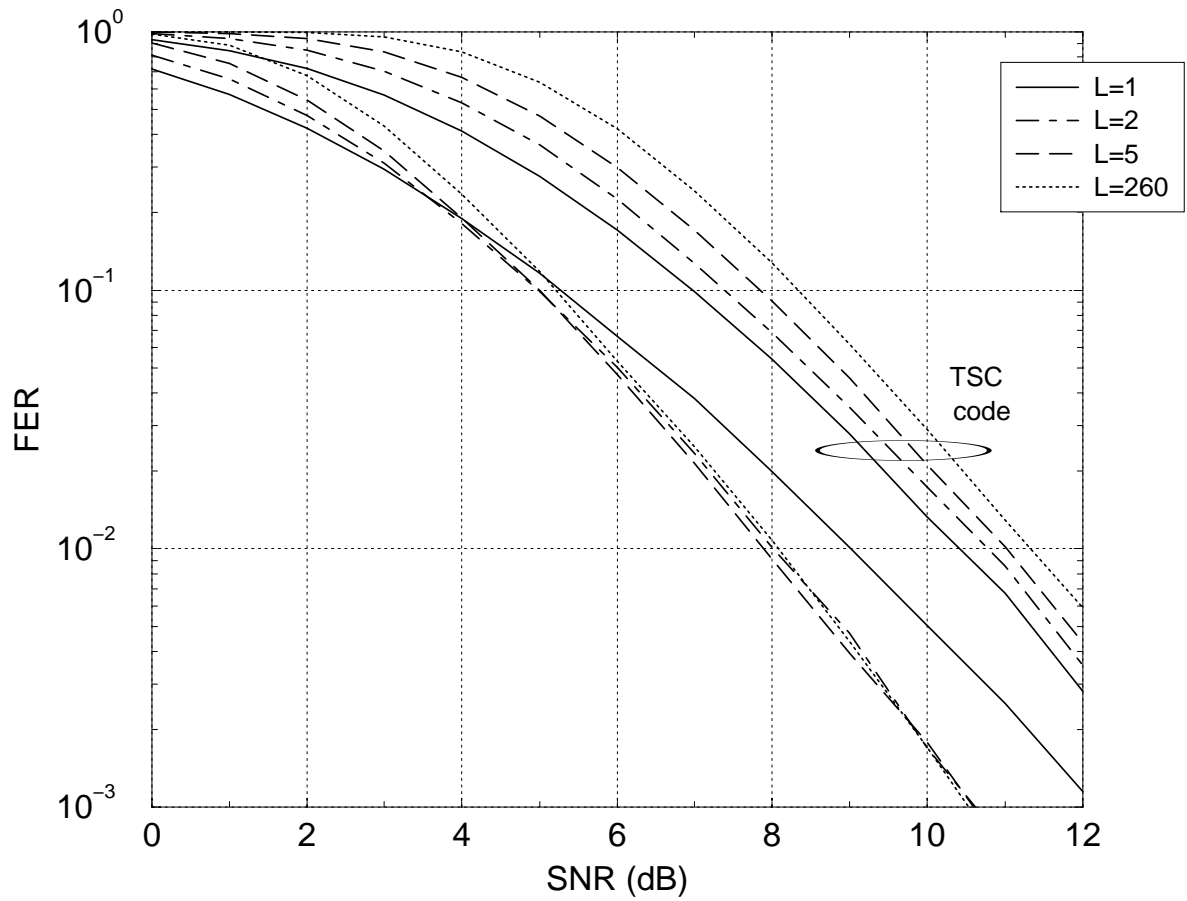}}
\caption{Comparison of our $4$ states QPSK MIMO(2,2), $2bps/Hz$, \ac{P-STC}
and the \ac{STC} in \cite{TarSesCal:98} (TSC) for different fading
rates.} \label{fig:QPSK_L_2x2_4stati_2bpsHz}
\end{figure}

\clearpage

%\begin{table}
%\caption{Best rate 1/2 \ac{P-STC} with \ac{BPSK}, $n=2$, $m=1$ on \ac{BFC} with $L=1$. Parameters with superscript $^{(1)}$ refer to the codes obtained with best convolutional codes for \ac{AWGN} channel. The performance factor is truncated with $d_H=2d_f^{(1)}-1$. Error performance refers to $E_b/N_0=20dB$ and $N=130$. }
%\center{
%\begin{tabular}{|c|c|c|c|c|c|c|c|c|}
  % after \\: \hline or \cline{col1-col2} \cline{col3-col4} ...
  %\hline
  %$\mu$ & {\bf Generators} &  $d_f$ & $\tilde{\eta}_{min}$ & $\tilde{F}_{\min}(m)/N$ & $d_f^{(1)}$ & $\tilde{F}_{\min}^{(1)}(m)/N$ & $\tilde{P}_{w\infty}$ (u.b.) & $FER$ \\
  %\hline
 %2 & $(1,2)_8$ & 2 &2 & 0.083 & 3 & 0.096 & $1.3\ 10^{-2}$ & $2.7\ 10^{-3}$\\
 %3 & $(3,4)_8$ & 3 &2 & 0.151 & 5 & 0.191 & $2.3\ 10^{-2}$ & $1.7\ 10^{-3}$\\
 %4 & $(13,15)_8$ & 6 &2 & 0.217 & 6 & 0.359 & $3.4\ 10^{-2}$ & $1.0\ 10^{-3}$\\
 %5 & $(23,31)_8$ & 6 &2 & 0.372 & 7 & 0.79 & $5.7\ 10^{-2}$ & $0.8\ 10^{-3}$\\
%\hline
%\end{tabular}
%}\label{tab:1/2,L=1,rx1}
%\end{table}

\begin{table}
\caption{Optimum rate $1/2$ \ac{P-STC} with \ac{BPSK}, $n=2$, $m=1$
on \ac{BFC} with $L=1$. Parameters with superscript $^{(1)}$ refer
to the codes obtained with best convolutional codes for \ac{AWGN}
channel. The performance factor is truncated with
$\delta_H=2d_f^{(1)}-1$. Error performance refers to $E_b/N_0=20dB$ and
$N=130$. } \center{
\begin{tabular}{|c|c|c|c|c|c|c|c|c|c|}
  % after \\: \hline or \cline{col1-col2} \cline{col3-col4} ...
  \hline
  $\mu$ & {\bf Generators} & {\bf Generators}$^{(1)}$  &  $d_f$ & $\tilde{\eta}_{min}$ & $\tilde{F}_{\min}(1)/N$ & $d_f^{(1)}$ & $\tilde{F}_{\min}^{(1)}(1)/N$ & $FER$ & $FER^{(1)}$ \\
  \hline
 2 & $(1,2)_8$ & $(1,3)_8$ & 2 &2 & 0.083 & 3 & 0.096 & $2.7\ 10^{-3}$ & $3.2\ 10^{-3}$\\
 3 & $(3,4)_8$ & $(5,7)_8$ & 3 &2 & 0.151 & 5 & 0.191 & $1.7\ 10^{-3}$ & $1.8\ 10^{-3}$\\
 4 & $(13,15)_8$ & $(15,17)_8$ & 6 &2 & 0.217 & 6 & 0.359 & $1.0\ 10^{-3}$ & $1.3\ 10^{-3}$\\
 5 & $(23,31)_8$ & $(23,35)_8$ & 6 &2 & 0.372 & 7 & 0.79 & $0.8\ 10^{-3}$ & $0.9\ 10^{-3}$\\
\hline
\end{tabular}
}\label{tab:1/2,L=1,rx1}
\end{table}

\begin{table}
\caption{Optimum rate 1/2 \ac{P-STC} with \ac{BPSK}, $n=2$, $m=1$ on
\ac{BFC} with $L=8$. Parameters with superscript $^{(1)}$ refer to
the codes obtained with best convolutional codes for \ac{AWGN}
channel. The performance factor is truncated with
$\delta_H=2d_f^{(1)}-1$. $\tilde{\eta}_{min}^{(R\ac{BFC})}$ is the diversity
achievable on the \ac{RBFC} with $nL$ fading blocks
\cite{ChiConTra:04}; the value of the Singleton bound is 9.}
\center{
\begin{tabular}{|c|c|c|c|c|c|c|}
  % after \\: \hline or \cline{col1-col2} \cline{col3-col4} ...
  \hline
  $\mu$ & {\bf Generators} &  $\tilde{\eta}_{min}$ & $\tilde{F}_{\min}(1)/N$ & $\tilde{\eta}_{min}^{(1)}$ & $\tilde{F}_{\min}^{(1)}(1)/N$ & $\tilde{\eta}_{min}^{(R\ac{BFC})}$ \\
  \hline
 2 & (1,3) & 2  & 0.031 & 2 & 0.031 &  4 \\
 3 &  (5,7) & 3  & 0.0039 & 3 & 0.0039 &  5 \\
 4 &  (07,15) & 4  & 9.8 $10^{-4}$ & 4 & 14.6 $10^{-4}$ &  6 \\
 5 &  (13,36) & 5  & 3.7 $10^{-4}$ & 5 &  7.5 $10^{-4}$    &  7 \\
 6 &  (57,75) & 6  & 2.3 $10^{-4}$ & 6 & 2.3 $10^{-4}$ &  8 \\
 7 &  (115,163) & 6  & 2.3 $10^{-5}$ & 6 & 1.0 $10^{-4}$ &  8 \\
\hline
\end{tabular}
}
\label{tab:1/2,L=8,rx1}
\end{table}

\begin{table}
\caption{Optimum \ac{P-STC} for a system with $n=2$, $m=1$ on
\ac{BFC} with $L=1$. Symbol $*$ indicates that the search based on
${F}_{\min}(\underline{c}_0,m)$ leads to the same code as the full
search.} \center{
\begin{tabular}{|c|c|c|c|c|c|c|c|c|}
  % after \\: \hline or \cline{col1-col2} \cline{col3-col4} ...
  \hline
  n & k & $\mu$ & h & {\bf Generators} &  $\tilde{\eta}_{min}$ & $\tilde{F}_{\min}(1)/N$ & ${F}_{\min}(\underline{c}_0,1)/N$ & $\tilde{F}_{\min}(2)/N$  \\
  \hline
 2 & 1 & 2 & 1 & (1,2) & 2  & 0.083 & 0.083 * &  0.0048 * \\
 2 & 1 & 3 & 1 & (3,4) & 2  & 0.15 & 0.16  &  0.017  \\
 2 & 1 & 4 & 1 & (13,15) & 2  & 0.22 & 0.27  &  0.011 *  \\
 4 & 1 & 2 & 2 & (1,2,3,1) & 2  & 0.082 & 0.087  &  0.003 * \\
 4 & 1 & 3 & 2 & (2,5,7,6) & 2  & 0.12 & 0.14  &  0.0011 * \\
 4 & 1 & 4 & 2 & (11,15,17,13) & 2  & 0.24 & 0.30  &  0.00083 * \\
 4 & 2 & 2 & 2 & (06,13,11,16) & 2  & 1.37 & 1.29 * &  0.073  \\
\hline
\end{tabular}
}
\label{tx2,L=1,rx1}
\end{table}

\begin{table}
\caption{Optimum \ac{P-STC} for a system with $n=3$, $m=1$ on
\ac{BFC} with $L=1$.  Symbol $*$ indicates that the search based
on ${F}_{\min}(\underline{c}_0,m)$  leads to the same code as the
full search. N.E.=Not evaluated.} \center{
\begin{tabular}{|c|c|c|c|c|c|c|c|}
  % after \\: \hline or \cline{col1-col2} \cline{col3-col4} ...
  \hline
  n & k & $\mu$ & h & {\bf Generators} &  $\tilde{\eta}_{min}$ & $\tilde{F}_{\min}(1)/N$ & ${F}_{\min}(\underline{c}_0,1)/N$  \\
  \hline
 3 & 1 & 2 & 1 & (1,2,3) & 2  & 0.026 & 0.021 *  \\
 3 & 1 & 3 & 1 & (2,3,4) & 3  & 0.030 & 0.033  \\
 3 & 1 & 4 & 1 & (11,12,15) & 3  & 0.033 & 0.044   \\
 6 & 1 & 2 & 2 & (1,1,2,2,3,3) & 2  & 0.027 & 0.021 * \\
 6 & 1 & 3 & 2 & (1,5,3,2,6,1) & 3  & 0.017 & 0.019   \\
 6 & 2 & 2 & 2 & (05,05,06,11,11,13) & 2  & N.E. & 0.05   \\
\hline
\end{tabular}
}
\label{tx3,L=1,rx1}
\end{table}

\begin{table}
\caption{Optimum \ac{P-STC} for a system with $n=4$, $m=1$ on
\ac{BFC} with $L=1$.  Symbol * indicates that the search based on
${F}_{\min}(\underline{c}_0,m)$  leads to the same code as the
full search. N.E.=Not evaluated.} \center{
\begin{tabular}{|c|c|c|c|c|c|c|c|}
  % after \\: \hline or \cline{col1-col2} \cline{col3-col4} ...
  \hline
  n & k & $\mu$ & h & {\bf Generators} &  $\tilde{\eta}_{min}$ & $\tilde{F}_{\min}(1)/N$ & ${F}_{\min}(\underline{c}_0,1)/N$  \\
  \hline
 4 & 1 & 2 & 1 & (1,1,2,3) & 2  & 0.016 & 0.013 *  \\
 4 & 1 & 3 & 1 & (1,3,5,7) & 3  & 0.0045 & 0.0039  \\
 4 & 1 & 4 & 1 & (03,05,11,16) & 4  &  N.E.   & 0.0057   \\
\hline
\end{tabular}
}
\label{tx4,L=1,rx1}
\end{table}
}

\end{document}